\documentclass[12pt, draftclsnofoot, onecolumn]{IEEEtran}
\usepackage{blindtext, graphicx,lipsum}
\usepackage[dvipsnames]{xcolor}
\usepackage{cuted,float}
\newlength\figureheight 
\newlength\figurewidth 
\usepackage{cite}
\usepackage{graphicx}
\usepackage{amsmath,amssymb,mathtools}
\usepackage{hyperref}
\usepackage{tkz-euclide}
\usepackage{multicol}
\usepackage{tikz}
\newlength\fheight
\newlength\fwidth
\usepackage{tkz-euclide}
\usepackage[utf8]{inputenc}
\usepackage{pgfplots} 
\usepackage{pgfgantt}
\usepackage{pdflscape}
\pgfplotsset{compat=newest} 
\pgfplotsset{plot coordinates/math parser=false}
\pgfplotsset{every  tick/.style={black,},ylabel style={font=\tiny},xlabel style={font=\tiny},tick label style={font=\tiny},legend style= {font=\scriptsize},
minor x tick num=1,minor y tick num=1,xminorticks=true,yminorticks=true,}

\def\endthebibliography{%
  \def\@noitemerr{\@latex@warning{Empty `thebibliography' environment}}%
  \endlist
}
\usepackage{amsthm}
\usetikzlibrary{arrows,plotmarks}
\usepackage{tikz-3dplot}
\usepackage{balance}
\newtheorem*{remark}{Remark}

\usepackage{algorithm}
\usepackage{algpseudocode}
\makeatletter
\renewcommand{\Function}[2]{%
  \csname ALG@cmd@\ALG@L @Function\endcsname{#1}{#2}%
  \def\jayden@currentfunction{#1}%
}
\newcommand{\funclabel}[1]{%
  \@bsphack
  \protected@write\@auxout{}{%
    \string\newlabel{#1}{{\jayden@currentfunction}{\thepage}}%
  }%
  \@esphack
}

\definecolor{cornellred}{rgb}{0.7, 0.11, 0.11}
\usepackage{longtable}
\usepackage{setspace}

\usepackage{caption,subcaption}
\newcommand{\ssymbol}[1]{^{\@fnsymbol{#1}}}

\begin{document}

\title{Hybrid Beamforming Design for Wideband mmWave Full-Duplex Systems}


\author{Elyes~Balti,~\IEEEmembership{Student Member,~IEEE,}
        and~Brian~L.~Evans,~\IEEEmembership{Fellow,~IEEE}
\thanks{E. Balti and B. L. Evans are with the Wireless Networking and Communications Group, Dept. Electrical and Computer Engineering, The University of Texas at Austin, Austin, TX 78712 USA (e-mails: ebalti@utexas.edu, bevans@ece.utexas.edu).}
}

\maketitle

\vspace*{-0.2in}

\begin{abstract}
\textcolor{black}{Recently, full duplex (FD) has been studied in 5G LTE millimeter wave (mmWave) cellular communications for New Radio in 3GPP releases 15-17. FD allows bidirectional transmission over the same resources and has the potential to reduce latency and double spectral efficiency. Self-interference (SI) is the primary drawback.  SI can be several orders of magnitude greater than the received signal power, saturate the analog-to-digital converters (ADCs) and degrade communication performance severely.
Massive mmWave antenna arrays may provide enough degrees of freedom for spatial multiplexing {\em and} SI suppression.
In this paper, we design spatial beamformers for the phased arrays already built into the FD basestation/relay to extend mmWave coverage to a single user. We propose alternating projections to design the precoder and combiner to maximize the sum of the uplink and downlink spectral efficiencies while bringing SI below the noise floor.  Our contributions include (1) hybrid analog/digital beamformer design algorithm to cancel SI in the analog domain to avoid ADC saturation and in the digital domain on each subcarrier; (2) full-digital beamformer design algorithm; and (3) analysis of spectral efficiency, energy efficiency and outage probability. In simulation, the proposed algorithms outperform beamsteering, singular value decomposition, angle search, and half-duplex techniques.}
\end{abstract}

\begin{IEEEkeywords}
Full-Duplex, Self-Interference, Single-User MIMO, MmWave Cellular, Hybrid Beamforming
\end{IEEEkeywords}
\IEEEpeerreviewmaketitle

\section{Introduction}
Over the last two decades, wireless networks have been evolving to meet the exponential growth in mobile data demand. In cellular networks, LTE standards have been increasing bandwidths, spectrum reuse, antennas, and modulation size accordingly. 4G LTE standards use frequency bands below 6 GHz and can provide up to 100 MHz in effective bandwidth depending on the deployment. 5G LTE standards not only continue using the 4G sub-6 GHz bands but also introduce millimeter wave (mmWave) frequency bands in the 10-300 GHz range\footnote[1]{Although a rigorous definition of mmWave frequencies would place them between 30 and 300 GHz, industry has loosely defined them to include the spectrum from 10 to 300 GHz.}. Designing beamforming (spatial filtering) algorithms for massive antenna arrays can overcome the high propagation losses in mmWave bands. 5G New Radio (NR) can achieve 10x increase in peak and average bit rates over 4G due to the large mmWave bandwidths, e.g. 700 MHz in the 24 GHz band and 850 MHz in the 28 GHz band. In Wi-Fi networks, the IEEE 802.11ad standard uses the unlicensed 57-64 GHz mmWave band to achieve high bit rates. \cite{heath,modular} 
\begin{figure}[t]
    \centering
    \includegraphics[scale=0.55]{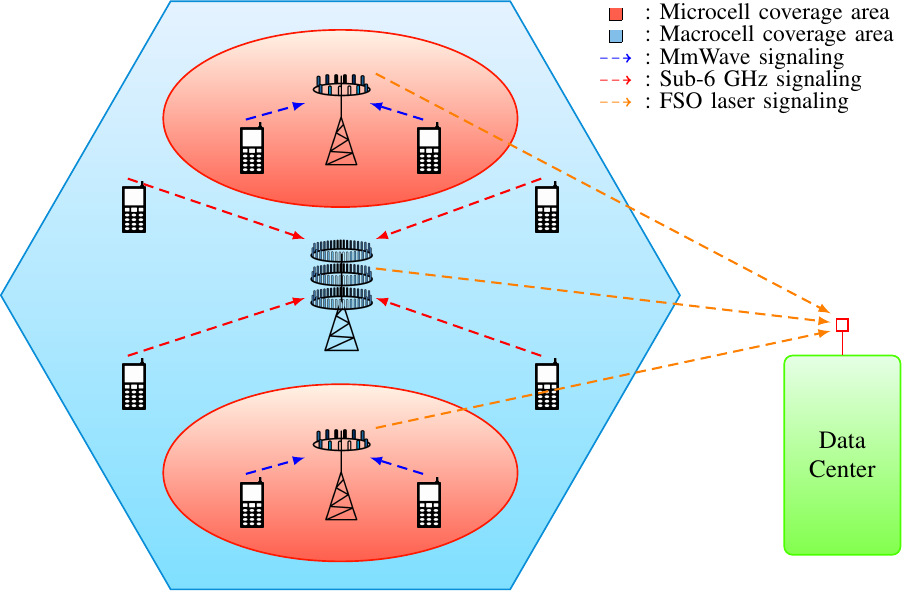}
    \caption{Outdoor heterogeneous mmWave cellular relay network with free space optics (FSO) backhaul wherein base stations are playing the key role of relays to improve coverage. Sub-6 GHz communications take place within macrocells (large area) where reliable links require high power to maintain coverage. In microcells, where cell area is small, mmWave signaling can reliably support high data rates to the users. \cite{j4} }
    \label{cellular}
\end{figure}

Full duplex (FD) could potentially reduce latency and double spectral efficiency in mmWave communication systems. These improvements make FD mmWave transceivers a potential candidate for applications such as platooning, advanced driving assistance system, autonomous driving, and vehicular clouds, which require huge bandwidth, high data rate and low latency.  Full duplex has been under study for New Radio by 3GPP Releases 15--17 for LTE standardization \cite{5GFD21}.

Fig. \ref{cellular} illustrates an example FD cellular system. Because FD systems transmit and receive using the same resource blocks, FD transceivers are subject to the near-far problem. The near-far problem can be illustrated using point-to-point communications. In system 1, the receiver receives a signal from its transmitter that could be several orders of magnitude stronger than the transmitter in system 2 due to propagation losses over the longer distance. The near-far problem can result in severe communication degradation.  In order to make FD systems practical, designing robust beamforming and interference cancellation techniques are critical.

\subsection{Self-Interference Cancellation Techniques}
\subsubsection{Antenna Array Architectures}
Interference cancellation can be realized using antenna separation, isolation, polarization \cite{1,2,3,4}, directional antennas \cite{5,6,7} or antenna placement to create null space at the receive array \cite{8,9}. Applicability of each technique depends on the hardware and other constraints. For example, passive SI cancellation using isolation and separation is limited for mobile devices due to their small size. Fortunately, interference suppression can be achieved in relaying systems because the transmit and receive arrays are not necessarily collocated. \cite{3} showed that directional arrays with a range of 4--6m of antenna isolation can achieve SI reduction as high as 80 dB. This extent of isolation can be applied in relaying systems; however, mobile devices cannot support such isolation due to their small size.

\subsubsection{Analog Circuitry}
This approach aims to suppress SI before the low noise amplifier (LNA) and analog-to-digital converter (ADC). The transmitted radio frequency (RF) signal is be extracted at the transmit power amplifier (PA) output, processed by an RF SI canceler and subtracted from the received signal. The analog RF SI canceler can be first applied \cite{6,7,10} to suppress only the internal coupling and reflections modeled by a programmable analog tapped-delay line (TDL) transversal filter. Adaptive digital RF cancellation can also be applied to suppress the SI components coming from the random external reflections by using a digital symbol-synchronous finite impulse-response (FIR) filter \cite{5,12,13}. 

 Traditionally, analog RF cancellation uses knowledge of the transmitted SI to cancel it before the receive LNA. A copy of the transmitted signal is obtained from the PA output and passed through a canceling circuit to reconstruct a copy of the received SI. The signal at the PA output includes the distortions of the transmitter (TX), which are reduced by the analog RF canceler.
 
The canceling circuit design is related to the nature of the SI channel. The  SI channel can be divided into internal reflections with a smaller number of paths, shorter delays and stronger amplitudes compared to the external (far-field) reflections. The internal reflections are static as they depend on the internal components and structure of the transceiver, while the external reflections vary according to the surrounding environment. Since it is difficult to adapt analog circuits to the variations in the external reflections, the analog RF canceler reduces only static internal reflections. The Renesas QHx220 chip \cite{5,13}, for example, takes the transmitted SI as input, changes its amplitude and phase to match the received SI, and subtracts the resulting signal from the received signal. This method achieves about 20 dB reduction in received SI \cite{13}.

\subsubsection{Digital Circuitry}
Processing the SI in the digital domain facilitates the use of adaptive filtering for a large number of reflected paths due to the external environment. The digital SI cancellation is based on the general transversal symbol-synchronous FIR structure where the constant tap-delay is equal to the signal sampling period and implemented as a D-flip flop clocked by the sampling clock. Here, only the tap-coefficients need to be specified from an estimate of the SI channel and thus we avoid the interaction between the delays and the attenuations as the case for the analog TDL. As a result, the digital processing can deal with a larger number of taps than the analog TDL to adapt to the varying external environment. 
The resulting canceling signal can be subtracted from the received signal at the RF input of the LNA/ADC to reduce the SI resulting from the external reflections further and keep the LNA/ADC from being overloaded. This operation requires an additional digital-to-analog (DAC) converter and an upconverting radio chain to generate the RF signal. The additional components will slightly change the generated SI leading to residual SI. This RF cancellation stage can provide 30 dB of SI cancellation \cite{10,12}, which, on top of the previously obtained 45 dB, still leaves a large amount of SI. 

The baseband cancellation stage represents the last line of defense against the SI by reducing it after the ADC. For this reason, we estimate the TX nonlinearities and residual SI channel resulting from the difference between the actual SI channel and equivalent channel generated by the previous cancellation stages. In addition, related work has proposed circuits for joint analog and digital SI cancellation. Digital SI cancellation is particularly suitable for MIMO systems as the cross-interference between antennas increases the number of taps needed to reduce the SI considerably.
In the same context, an all-digital SI cancellation based on a new FD transceiver structure significantly reduces transceiver impairments \cite{18}. This technique consists of an intermediary receiver (RX) chain to obtain a digital replica of the transmitted SI signal that will be used to cancel the SI signal and TX imperfections. A combination of this digital technique and passive RF cancellation significantly reduces the SI to be 3 dB higher than the noise floor, thereby resulting in 67-76$\%$ rate enhancement compared to the conventional half duplex (HD) systems operating at 20dBm of transmit power \cite{j1}.

\subsubsection{Spatial Beamforming}
Another technique, spatial suppression, has been effective in mitigating SI for FD MIMO systems \cite{19,20}. This technique leverages available degrees of freedom (DoF) from the multiple antennas to cancel SI while maintaining acceptable multiplexing gain. This approach does not require any additional analog circuitry. Analog beamforming architectures are cost-efficient because they only require phase shifters. The CA constraint of the phase shifters reduces the DoF and communications performance.
Similar work \cite{21} proposed a design to find the optimal analog beamformers.  When projecting the solution onto the subspace of the CA constraint, the interference cancellation constraint is violated, which causes significant performance losses. In addition, related work proposed hybrid beamforming designs for FD systems. \cite{hbd} proposed an iterative optimization algorithm to minimize the SI power and improve the spectral efficiency while maintaining a reasonable number of iterations for convergence. \cite{hbd1} proposed a beamforming design algorithm to maximize the spectral efficiency for a dual-hop FD relaying system wherein the performance is compared to HD and upper bound as benchmarks. 

\subsection{Contributions}
We consider a wideband mmWave cellular system in which an FD base station (BS) independently communicates with an uplink UE and a downlink UE.  The goal is to design hybrid beamformers to cancel the loopback SI, avoid ADC saturation, \textcolor{black}{beat the HD mode} and provide the uplink UE, which is vulnerable to the SI, with acceptable spectral efficiency relative to the downlink UE which is interference-free. The contributions follow:
\begin{itemize}
   \item \textcolor{black}{We aim to minimize the SI power by jointly designing the full-digital beamformers for the BS and the UEs by applying the zero-forcing max power algorithm based on the idealized simplifying assumption of having perfect channel state information (CSI).} 
    \item \textcolor{black}{We propose to design the hybrid analog/digital beamformers at the BSs and UEs in two stages. The first stage designs the analog beamformers using \emph{Alternating Projections} (successive projections between the zero-forcing null-space and the constant amplitude subspace) to project the beamformers on the optimal subspace to wipe out the SI in the analog domain and avoid ADC saturation. In the second stage, we derive a closed-form solution for the digital beamformers to cancel the residual SI on each subcarrier. Given the uplink UE is corrupted by SI, with sufficient number of iterations, the hybrid beamforming algorithm eliminates the SI power and nearly approaches the interference-free downlink UE performance.}
    \item We present quantitative comparisons of our proposed algorithms through simulation, wherein our proposed design achieves better performance gains than beam steering, Singular Value Decomposition (SVD), angle search techniques and \textcolor{black}{HD mode}.
     \item We analyze spectral efficiency, energy efficiency, outage probability \textcolor{black}{and gain/loss of spectral efficiency} for the proposed hybrid beamformer. For spectral efficiency, we provide the full-digital beamformer and upper bound as benchmarks.
\end{itemize}

\subsection{Organization}
The rest of the paper is organized as follows: Section 2 discusses the system and channel models, while the beamforming design for full-digital and hybrid architectures are detailed in Section 3. Energy efficiency and outage probability are analyzed in Section 4, whereas numerical results and their discussions appear in Section 5. Section 6 concludes the paper.

\subsection{Notation}
\textcolor{black}{In this paper, bold lowercase $\mathbf{x}$ denotes column vectors, bold uppercase $\mathbf{X}$ denotes matrices, non-bold letters $x, X$ denote scalar values, and calligraphic letters $\mathcal{X}$ denote sets. Using this notation, $|x|$ is the absolute value of a scalar, $\|\mathbf{x}\|_2$ is the $\ell_2$ norm, $\|\mathbf{x}\|_0$ is the $\ell_0$ norm, $\| \mathbf{X}\|_F$ is the Frobenius norm, $\sigma_n(\mathbf{X})$ is the $n$-th singular value of $\mathbf{X}$ in decreasing order, $\mathrm{det}(\mathbf{X})$ denotes the determinant, $\mathrm{Tr}(\mathbf{X})$ denotes the trace, $\mathbf{X}^*$ is the Hermitian or conjugate transpose, $\mathbf{X}^T$ is the matrix transpose, $\mathbf{X}^{-1}$ denotes the inverse of a square non-singular matrix, $\mathbf{X}^{\dagger}$ denotes the pseudo-inverse, $[\mathbf{x}]_n$ is the $n$-th entry of $\mathbf{x}$, $|\mathcal{X}|$ is the cardinality of set $\mathcal{X}$. We use the notation $\mathcal{CN}(\mathbf{m},\mathbf{\Sigma})$ to denote a complex circularly symmetric Gaussian random vector with mean $\mathbf{m}$ and covariance $\mathbf{\Sigma}$. We use $\mathbb{E}[\cdot]$ to denote the expectation and $\mathbb{P}[\cdot]$ is the probability measure.}


\begin{figure}[tbhp]
    \centering
    \includegraphics[width=\linewidth]{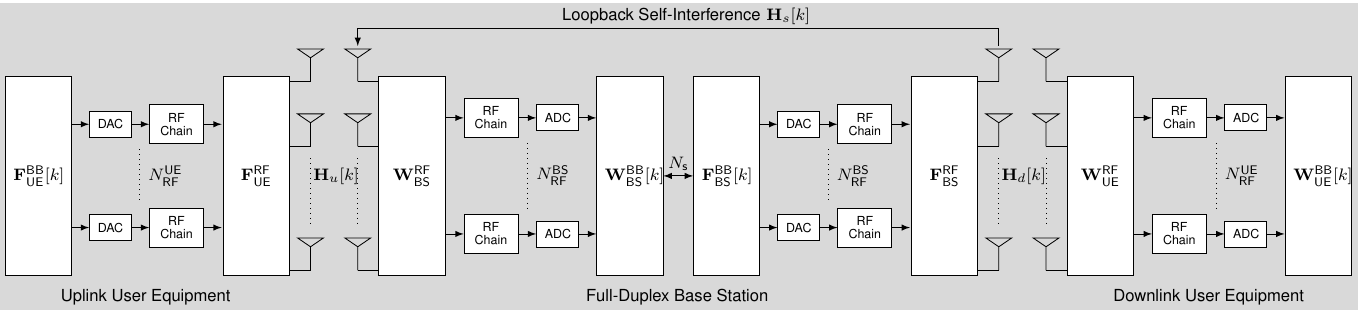}
    \caption{Hybrid architecture of dual-hop full duplex (FD) relaying system. The uplink user equipment (UE) sends data to the base station (BS) independently from the data intended for the downlink UE sent by the BS. Self-interference (SI) is due to the BS transmitting and receiving simultaneously using the same resource blocks, and we model the SI leakage in the time domain by the wideband channel matrix denoted $\mathsf{H}_s[\ell]$.}
    \label{system}
\end{figure}

\section{System Model}
Fig.\ \ref{system} shows the wideband mmWave full-duplex system with hybrid analog/digital beamforming. Transmission uses OFDM signaling with $K$ subcarriers. At the $k$-th subcarrier, symbols $s[k]$ are transformed to the time domain using a $K$-point IDFT. A cyclic prefix (CP) of length ($L_c$) is appended to the time domain samples before applying the precoder. The OFDM block is formed by the CP followed by the $K$ time domain samples and the data symbols follows $\mathbb{E}[\mathbf{s}[k]\mathbf{s}^*[k]]=\frac{\rho}{KN_{\mathsf{s}}}\mathbf{I}$, where $\rho$ is the total average transmit power for the data i.e., without considering the CP, per OFDM symbol. We assume the maximum delay spread of the channel is within the CP duration.
This description applies equally for uplink and downlink transmission.

For uplink, the received signal at the BS in the $k$-th subcarrier is given by
\begin{equation}
\mathbf{y}_{\mathsf{uplink}}[k] = \underbrace{\sqrt{\rho_u}\mathbf{W}^*_{\mathsf{BS}}[k]\mathbf{H}_u[k]\mathbf{F}_{\mathsf{UE}}[k]\mathbf{s}_u[k]}_{\textsf{Desired Signal}} +
\underbrace{\sqrt{\rho_s} \mathbf{W}_{\mathsf{BS}}^*[k]\mathbf{H}_{s}[k]\mathbf{F}_{\mathsf{BS}}[k]\mathbf{s}_d[k]}_{\textsf{Self-Interference}} +
\underbrace{\mathbf{W}^*_{\mathsf{BS}}[k] \mathbf{n}_{\mathsf{BS}}[k]}_{\textsf{AWGN}}   
\end{equation}
where $\mathbf{W}_{\mathsf{BS}}[k] \in \mathbb{C}^{N_{\mathsf{BS}} \times N_{\mathsf{s}}}$, 
$\mathbf{F}_{\mathsf{BS}}[k] \in \mathbb{C}^{N_{\mathsf{BS}} \times N_{\mathsf{s}}}$ and
$\mathbf{F}_{\mathsf{UE}}[k] \in \mathbb{C}^{N_{\mathsf{UE}} \times N_{\mathsf{s}}}$ 
are the full-digital combiner and precoder at the BS and full-digital precoder at the uplink UE, respectively, for the $k$-th subcarrier.
$\mathbf{H}_u[k] \in \mathbb{C}^{N_{\mathsf{BS}} \times N_{\mathsf{UE}}}$ 
and 
$\mathbf{H}_s[k] \in \mathbb{C}^{N_{\mathsf{BS}} \times N_{\mathsf{BS}}}$ 
are the uplink and SI, respectively, for the $k$-th subcarrier, while $\mathbf{s}_u[k]$, $\mathbf{s}_d[k]$ and $\mathbf{n}_{\mathsf{BS}}[k]$ are the UE data sent to BS, BS data sent to downlink UE, and additive white Gaussian noise (AWGN) at the BS with $\mathcal{CN}(0,\sigma_u^2)$, respectively; $\rho_u$ and $\rho_s$ are the average BS transmit power and SI power, respectively.


For downlink, the received signal at the UE at the $k$-th subcarrier is expressed by
\begin{equation}
\mathbf{y}_{\mathsf{downlink}}[k] = \sqrt{\rho_d}\mathbf{W}_{\mathsf{UE}}^*[k]\mathbf{H}_d[k]\mathbf{F}_{\mathsf{BS}}[k]\mathbf{s}_d[k] + \mathbf{W}_{\mathsf{UE}}^*[k] \mathbf{n}_{\mathsf{UE}}[k]
\end{equation}
$\mathbf{W}_{\mathsf{UE}}[k] \in \mathbb{C}^{N_{\mathsf{UE}} \times N_{\mathsf{s}}}$ 
is the fully digital combiner at downlink UE, $\rho_d$ is BS power received by UE, $\mathbf{n}_{\mathsf{UE}}[k]$ is AWGN at UE with $\mathcal{CN}(0,\sigma_d^2)$ and 
$\mathbf{H}_d[k]\in \mathbb{C}^{N_{\mathsf{UE}} \times N_{\mathsf{BS}}}$ is $k$-th downlink subcarrier.
Unlike the downlink case, the uplink received signal is corrupted by loopback SI at the FD BS.



\subsection{Baseband Channel Model}
In this work, we assume uplink and downlink MIMO channels are wideband, with a delay tap length $L$ in the time domain. The $\ell$-th delay tap of the channel is represented by an $N_{\mathsf{RX}} \times N_{\mathsf{TX}}$ matrix, $\ell = 0,\ldots,L-1$, which assuming a geometric cluster and ray based channel model \cite{anum} 
\begin{equation}\label{channel}
\begin{split}
\mathsf{H}[\ell] = \sqrt{\frac{N_{\mathsf{RX}}N_{\mathsf{TX}}}{\gamma}} \sum_{c=0}^{C-1}\sum_{r_c=0}^{R_c-1} \alpha_{r_c} p(\ell T_s - \tau_{c}-\tau_{r_c}) \mathbf{a}_{\mathsf{RX}}(\theta_c+\vartheta_{r_c})   \mathbf{a}_{\mathsf{TX}}^*(\phi_c+\varphi_{r_c})    
\end{split}
\end{equation}
where $C$ and $R_c$ are the numbers of clusters and rays per cluster; $T_s$ is the signaling interval; $\tau_c$ is the cluster mean time delay; and $\theta_c$ and $\phi_c$ are angles of arrival (AoA) and departure (AoD). Each ray has a relative time delay $\tau_{r_c}$, relative AoA ($\vartheta_{r_c}$) and AoD ($\varphi_{r_c}$) shifts, and complex gain $\alpha_{r_c}$. Here, $\gamma$ is the pathloss and $p(\tau)$ is the raised cosine pulse shape evaluated at $\tau$. Also, $\mathbf{a}_{\mathsf{RX}}(\theta)$ and $\mathbf{a}_{\mathsf{TX}}(\phi)$ are the RX and TX antenna array response vectors, respectively, and
\begin{equation}
\mathbf{a}_{\mathsf{RX}}(\theta) = \frac{1}{\sqrt{N_{\mathsf{RX}}}}\left[1,e^{j\frac{2\pi d}{\lambda} \sin(\theta)},\ldots,e^{j\frac{2\pi d}{\lambda}\left(N_{\mathsf{RX}} -1\right)\sin(\theta)}  \right]^{T}    
\end{equation}
The channel at the $k$-th subcarrier, where $k = 0, 1, \ldots K-1$, is given by
\begin{equation}
\mathbf{H}[k] = \sum_{\ell=0}^{L-1} \mathsf{H}[\ell] e^{-j\frac{2\pi k}{K}\ell}   
\end{equation}

\subsection{Self-Interference Channel Model}
\begin{figure}[b]
    \centering
    \includegraphics{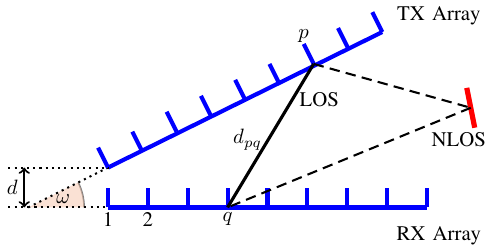}
    \caption{Relative positions of TX and RX arrays collocated at a full-duplex basestation. For the line-of-sight channel, due to collocation, a far-field assumption that the SI impinges on the RX array as a planar wave does not hold; instead, the SI impinges as a spherical wave. \cite{bm20}}
    \label{position}
\end{figure}

Per Fig.~$\ref{position}$, SI leakage at the BS is modeled by channel matrix $\mathsf{H}_s$. The SI channel is decomposed into a line-of-sight (LOS) component modeled by $\mathsf{H}_{\mathsf{los}}$ and a non-LOS (NLOS) leakage described by $\mathsf{H}_{\mathsf{nlos}}$ which follows the channel model (\ref{channel}). The LOS SI leakage matrix
\begin{equation}\label{eq2.2}
 [\mathsf{H}_{\mathsf{los}}]_{pq} = \frac{1}{d_{pq}}e^{-j2\pi\frac{d_{pq}}{\lambda}}    
\end{equation}
where $d_{pq}$ is the distance between the $p$-th TX antenna and $q$-th RX antenna at BS. The aggregate $\ell$-th tap SI channel $\mathsf{H}_s[\ell]$ can be expressed as \cite[Eq.~(2)]{si} where $\kappa$ is the Rician factor:
\begin{equation}\label{eq2.3}
\mathsf{H}_s[\ell] = \underbrace{\sqrt{\frac{\kappa}{\kappa+1}}\mathsf{H}_{\mathsf{los}}}_{\mathsf{Near-Field}} + \underbrace{\sqrt{\frac{1}{\kappa+1}}\mathsf{H}_{\mathsf{nlos}}[\ell]}_{\mathsf{Far-Field}}    
\end{equation}

\subsection{Transceiver Impairments}
5G mmWave cellular systems operate in the 24, 28, 37, 39 and 47 GHz bands. In US FCC auctions, the corresponding transmission bandwidths are 0.7, 0.85, 1, 1, and 1.4 GHz broken into 100 MHz subbands.  MmWave systems need very large antenna arrays to overcome high propagation losses. The massive jump in the number of antenna elements, transmission bandwidth, and data converter conversion rates for 5G mmWave systems has lead to new basestation architectures for energy efficiency, such as hybrid analog/digital beamforming.

Introducing FD transceivers also has several design challenges to achieve doubling spectral efficiency. High power amplifier (HPA) nonlinearity can severely degrade the system performance such as the creation of irreducible outage/error floor and/or spectral efficiency saturation. In addition, the amplifier produces intermodulation products that translate into spectral regrowth (or spectral shoulder) resulting in interference in adjacent subcarriers and loss of information.

Conventional techniques to compensate nonlinear effects include those based on Bussgang Linearization. Moreover, heuristic HPA nonlinearity models have been proposed such as soft envelope limiter, traveling wave tube amplifier, and solid state power amplifier \cite{j4,j2,c4}.

Furthermore, mmWave transmitter impairments may affect SI cancellation for FD systems. In order to subtract the received SI signal, any modifications that occurs on the transmitter side would have to be captured. This includes channel features and analog components such as the PA and IQ mixer. The transmitted SI is slightly modified as it moves through the transmit chain and such modifications are negligible compared to the desired signal; however, they are of significant magnitude compared to the intended signal and will limit the performance of FD systems. 

The inband image resulting from the transmit IQ mixer is about 30 dB lower than the direct signal. In the presence of strong SI of 50 dB higher than the transmit signal, the IQ image causes additional interference to the intended signal and also has to be reduced. Previous work selects FD transceiver component impairments \cite{h4,h25,h35}. Alternate high-speed DACs directly convert the baseband signal to RF, which can avoid many nonlinear distortions related to upconversion.
\begin{remark}
Nonlinearities and other impairments in mmWave analog/RF circuits degrade FD communication performance. Although modeling these impairments is out of scope for this work, we include them as additional sources of SI. The aggregate SI power used in this work is about 120 dB. The near-far problem incurs SI of about 20--60 dB (depending on the UE being near the BS, at mid-range or at cell edge) and the remaining SI comes from transceiver impairments.
\end{remark}
\section{Beamforming Design}
\subsection{Full-Digital Beamforming}
\label{fulldigitialbeamforming}
In the full-digital domain, the uplink spectral efficiency is
\begin{equation}{\label{uplinkratedigital}}
\begin{split}
\mathcal{I}(\rho_u) = \frac{1}{K}\sum_{k=0}^{K-1}\log\det\left(\mathbf{I}_{N_{\mathsf{s}}} + \rho_u\mathbf{W}^*_{\mathsf{BS}}[k] \mathbf{H}_u[k]\mathbf{F}_{\mathsf{UE}}[k]\mathbf{Q}_u[k]^{-1}\mathbf{F}^*_{\mathsf{UE}}[k] \mathbf{H}^*_u[k]\mathbf{W}_{\mathsf{BS}}[k]     \right)  
\end{split}
\end{equation}
Here $\mathbf{Q}_u[k]$ is the SI plus noise covariance matrix at the $k$-th subcarrier given by
\begin{equation}
\begin{split}
\mathbf{Q}_u[k] =\rho_s  \mathbf{W}^*_{\mathsf{BS}}[k] \mathbf{H}_s[k]  \mathbf{F}_{\mathsf{UE}}[k]\mathbf{F}^*_{\mathsf{UE}}[k] \mathbf{H}_s^*[k]\mathbf{W}_{\mathsf{BS}}[k]+\sigma_u^2  \mathbf{W}^*_{\mathsf{BS}}[k]\mathbf{W}_{\mathsf{BS}}[k].    
\end{split}
\end{equation}
where 
$\sigma_u^2 = -173.8 \, {\rm dB} + 10 \log_{10} ({\rm Bandwidth})$ \cite{bm20}.

For downlink scenario, the spectral efficiency is given by
\begin{equation}
\begin{split}
\mathcal{I}(\rho_d) = \frac{1}{K}\sum_{k=0}^{K-1}\log\det\left(\mathbf{I}_{N_{\mathsf{s}}} + \rho_d\mathbf{W}^*_{\mathsf{UE}}[k] \mathbf{H}_d[k]\mathbf{F}_{\mathsf{BS}}[k]\mathbf{Q}_d[k]^{-1}\mathbf{F}^*_{\mathsf{BS}}[k] \mathbf{H}^*_d[k]\mathbf{W}_{\mathsf{UE}}[k]     \right)     \end{split}   
\end{equation}
where $\mathbf{Q}_d[k]$ is the noise covariance matrix for the $k$-th subcarrier.

The design objective is to construct beamformers that are robust to SI and provide the uplink UE with acceptable spectral efficiency compared to the downlink UE, given that the BS reception of the former is affected by SI. We decompose the beamforming design into two phases. The first stage maximizes the rate for the uplink user, and the downlink user beamforming is designed in the second stage. Starting with the uplink scenario, we adopt a sub-optimal approach in which the zero-forcing constraint is imposed, thereby leading to the following optimization problem:
\begin{equation}
\max_{\substack{\mathbf{W}_{\mathsf{BS}}[k],\mathbf{F}_{\mathsf{BS}}[k],\mathbf{F}_{\mathsf{UE}}[k],\mathbf{W}_{\mathsf{UE}}[k]\\k=0,\ldots,K-1}} \mathcal{I}(\rho_u) + \mathcal{I}(\rho_d) 
\end{equation}
\begin{equation*}
\begin{split}
\textsf{Subject to}~&\mathbf{W}^*_{\mathsf{BS}}[k]\mathbf{W}_{\mathsf{BS}}[k] = \mathbf{I}_{N_{\mathsf{s}}} \\&
\mathbf{F}^*_{\mathsf{BS}}[k]\mathbf{F}_{\mathsf{BS}}[k] = \mathbf{I}_{N_{\mathsf{s}}} \\&
\mathbf{F}^*_{\mathsf{UE}}[k]\mathbf{F}_{\mathsf{UE}}[k] = \mathbf{I}_{N_{\mathsf{s}}} \\&
\mathbf{W}^*_{\mathsf{UE}}[k]\mathbf{W}_{\mathsf{UE}}[k] = \mathbf{I}_{N_{\mathsf{s}}} \\&
\mathbf{W}^*_{\mathsf{BS}}[k]\mathbf{H}_s[k]\mathbf{F}_{\mathsf{BS}}[k] = \mathbf{0}~(\textsf{\scriptsize{Zero-Forcing Constraint}})\\&
k = 0,\ldots, K-1
\end{split}
\end{equation*}

We observe that the BS beamformers have to be jointly designed to maximize the beamformed received power and minimize the SI power simultaneously. To cope with the coupling in the uplink and downlink optimization, we fix the precoder and solve for the combiner, and then fix the combiner and solve for the precoder. The ZF cyclic maximization process will iterate to maximize the cost function. 
Convergence is guaranteed since the cost function is bounded as the beamformers are normalized and transmit power $\rho$ is constrained. 
The optimization problem is
\begin{equation}
    \max_{\mathbf{X}}~\log\det\left( \mathbf{I}_N + \rho\mathbf{X}^*\mathbf{A}\mathbf{A}^*\mathbf{X}\right)
\end{equation}
\begin{equation*}
\begin{split}
\textsf{Subject to}~&\mathbf{X}^*\mathbf{X} = \mathbf{I}_N\\&
\mathbf{X}^*\mathbf{C} = \mathbf{0}
\end{split}
\end{equation*}
where $\mathbf{X} \in \mathbb{C}^{M\times N}$ is the beamformer, $\mathbf{A}\in \mathbb{C}^{M\times N}$ is a beamformed subcarrier and $\mathbf{C}\in \mathbb{C}^{M\times P}$ is a beamformed SI channel that spans the SI subspace.
This generic form is for a single subcarrier; the design algorithm will be repeated for all subcarriers. \textcolor{black}{Algorithm \ref{full-digital-beamforming} solves this problem and includes a computational complexity analysis. As terms go to infinity, the overall complexity of
$K N_{\mathsf{outer}} \left( 6 \, N_{\mathsf{s}} N_{\mathsf{BS}}^2 +
     18 \, N_{\mathsf{BS}} N_{\mathsf{s}}^2 +
     18 \, N_{\mathsf{UE}} N_{\mathsf{s}}^2 +
      4 \, N_{\mathsf{BS}}N_{\mathsf{s}}N_{\mathsf{UE}} \right)$
is dominated by $K N_{\mathsf{outer}} N_{\mathsf{s}} N_{\mathsf{BS}}^2$.} 
\textcolor{black}{\begin{proof}
The proof of the full-digital solution is provided by Appendix A.
\end{proof}}

\begin{remark}
Given the digital beamformer solution $\mathbf{X} \in \mathbb{C}^{M\times N}$, $M$ and $N$ are the numbers of antennas and spatial streams, respectively. $M$ should be large enough to sustain $N$ spatial streams and the remaining $P=M-N$ degrees of freedom should be used to cancel SI.
\end{remark}



\begin{algorithm}
\caption{Full-Digital Beamforming 
    \hfill \textcolor{blue}{Complex Multiplications for Highest-Order Terms}}
\label{full-digital-beamforming}
\begin{algorithmic}[1]
\Function{Baseband}{$\mathbf{A},\mathbf{C},N$} \funclabel{alg:a} \label{alg:a-line}
        \hfill \textcolor{blue}{
        $\mathbf{A},\mathbf{X}\in\mathbb{C}^{M\times N};
        \mathbf{C}\in\mathbb{C}^{M\times P};
        \mathbf{P}_{\bot} \in\mathbb{C}^{M\times M};
        M > N$}
    \State $\mathbf{P}_{\bot} \gets \mathbf{I}-\mathbf{C}\mathbf{C}^{*}$
        \hfill \textcolor{blue}{$P M^2$}
    \State Compute $\mathbf{P}_\bot\mathbf{A}$
        \hfill \textcolor{blue}{$N M^2$}
    \State $\mathbf{X} \gets N~\text{Dominant left singular vectors \cite{svdComplexityAnalysis} of } \mathbf{P}_\bot\mathbf{A}$
        \hfill \textcolor{blue}{$9 \, M N^2$}
    \State \Return $\mathbf{X}$
\EndFunction
    \hfill \textcolor{blue}{$9 \, M N^2 + (N+P) M^2$}
\Statex
\State \textbf{Input} $\mathbf{H}_s[k], \mathbf{H}_u[k] ,\mathbf{H}_d[k],~k=0,\ldots,K-1$   
    \hfill
    \textcolor{blue}{
    $\mathbf{H}_s[k] \in \mathbb{C}^{N_{\mathsf{BS}} \times N_{\mathsf{BS}}}; \mathbf{H}_u[k], \mathbf{H}^{*}_d[k] \in \mathbb{C}^{N_{\mathsf{BS}} \times N_{\mathsf{UE}}}$}
\State \textbf{Initialize} $\mathbf{F}_{\mathsf{BS}}[k], \mathbf{F}_{\mathsf{UE}}[k],~k=0,\ldots,K-1$
    \hfill
    \textcolor{blue}{
    $\mathbf{F}_{\mathsf{BS}}[k] \in \mathbb{C}^{N_{\mathsf{BS}} \times N_{\mathsf{s}}}; \mathbf{F}_{\mathsf{UE}}[k] \in \mathbb{C}^{N_{\mathsf{UE}} \times N_{\mathsf{s}}}$}
\For{$k\gets 0:K-1$}
\For{$t\gets 1:N_{\mathsf{outer}}$}
\State Compute $\mathbf{H}_u[k]\mathbf{F}_{\mathsf{UE}}[k]$
    \hfill \textcolor{blue}{$N_{\mathsf{BS}}N_{\mathsf{s}}N_{\mathsf{UE}}$}
\State Compute $\mathbf{H}_s[k]\mathbf{F}_{\mathsf{BS}}[k]$
    \hfill \textcolor{blue}{$N_{\mathsf{s}} N^2_{\mathsf{BS}}$}
\State $\mathbf{W}_{\mathsf{BS}}[k] \gets$ {\textsc{Baseband}}$\left(\mathbf{H}_u[k]\mathbf{F}_{\mathsf{UE}}[k],\mathbf{H}_s[k]\mathbf{F}_{\mathsf{BS}}[k], N_{\mathsf{s}}\right)$
    \hfill \textcolor{blue}{$9 \, N_{\mathsf{BS}} N_{\mathsf{s}}^2 + 2 N_{\mathsf{s}} N^2_{\mathsf{BS}}$}
\State Compute $\mathbf{H}^*_u[k]\mathbf{W}_{\mathsf{BS}}[k]$
    \hfill \textcolor{blue}{$N_{\mathsf{BS}}N_{\mathsf{s}}N_{\mathsf{UE}}$}
\State $\mathbf{F}_{\mathsf{UE}}[k] \gets N_{\mathsf{s}}$~\text{Dominant left singular vectors of} $\mathbf{H}^*_u[k]\mathbf{W}_{\mathsf{BS}}[k]$ 
    \hfill \textcolor{blue}{$9 \, N_{\mathsf{UE}} N_{\mathsf{s}}^2$}
\State Compute $\mathbf{H}_d[k]\mathbf{F}_{\mathsf{BS}}[k]$ 
    \hfill \textcolor{blue}{$N_{\mathsf{BS}} N_{\mathsf{s}} N_{\mathsf{UE}}$}
\State $\mathbf{W}_{\mathsf{UE}}[k] \gets N_{\mathsf{s}}$~\text{Dominant left singular vectors of} $\mathbf{H}_d[k]\mathbf{F}_{\mathsf{BS}}[k]$ 
    \hfill \textcolor{blue}{$9 \, N_{\mathsf{UE}} N_{\mathsf{s}}^2$}
\State Compute $\mathbf{H}^*_d[k]\mathbf{W}_{\mathsf{UE}}[k]$
    \hfill \textcolor{blue}{$N_{\mathsf{BS}} N_{\mathsf{s}} N_{\mathsf{UE}}$}
\State Compute $\mathbf{H}_s^*[k]\mathbf{W}_{\mathsf{BS}}[k]$
    \hfill \textcolor{blue}{$N_{\mathsf{s}} N_{\mathsf{BS}}^2$}
\State $\mathbf{F}_{\mathsf{BS}}[k] \gets$ {\textsc{Baseband}}$\left(\mathbf{H}^*_d[k]\mathbf{W}_{\mathsf{UE}}[k],\mathbf{H}_s^*[k]\mathbf{W}_{\mathsf{BS}}[k],N_{\mathsf{s}}\right)$
    \hfill \textcolor{blue}{$9 \, N_{\mathsf{BS}} N_{\mathsf{s}}^2 + 2 N_{\mathsf{s}} N^2_{\mathsf{BS}}$}
\EndFor
\EndFor
    \hfill \textcolor{blue}{Overall:
    $ K N_{\mathsf{outer}} \left( 6 \, N_{\mathsf{s}} N_{\mathsf{BS}}^2 +
     18 \, N_{\mathsf{BS}} N_{\mathsf{s}}^2 +
     18 \, N_{\mathsf{UE}} N_{\mathsf{s}}^2 +
      4 \, N_{\mathsf{BS}}N_{\mathsf{s}}N_{\mathsf{UE}} \right)$}
\State \Return $\mathbf{F}_{\mathsf{BS}}[k],\mathbf{W}_{\mathsf{BS}}[k],\mathbf{W}_{\mathsf{UE}}[k],\mathbf{F}_{\mathsf{UE}}[k],~k=0,\ldots,K-1$
\end{algorithmic}
\end{algorithm}

\textcolor{black}{\subsection{Alternating Projections Method} \label{altprojsec}
This method computes a point at the intersection of convex sets by using a sequence of projections onto the sets. Although the method progresses slowly, it is particularly useful when we have an efficient procedure for carrying out the projections such as by an analytical formula.}

\textcolor{black}{We use only Euclidean norm, distance, and projection. Suppose $A$ and $B$ are closed convex sets in $\mathbf{R}^{n}$, and let $P_A$ and $P_B$ denote the projection on $A$ and $B$, respectively. The algorithm starts with any $x_0 \in A$, and then alternately projects onto $A$ and $B$:
\begin{equation}
    y_k = P_B(x_k),~~x_{k+1} = P_A(y_k),~~k=0,1,2,\ldots
\end{equation}
This generates a sequence of points $x_k\in A$ and $y_k\in B$. A basic result \cite{Cheney} is the following. If $A \cap B = \emptyset$, then the sequences $x_k$ and $y_k$ both converge to a point $x^*\in A\cap B$.}

\begin{figure}[tbhp]
\begin{subfigure}[b]{0.5\textwidth}
\centering
\setlength\fheight{3cm}
\setlength\fwidth{7cm}
\begin{tikzpicture}

    \draw[rotate=45,fill=none] (0,0) ellipse (30pt and 65pt);

    \draw[rotate=-35,fill=none] (1.5,2) ellipse (30pt and 65pt);
    \draw plot [mark=*, mark size=2] coordinates{(1,0.45)}; 
    \draw plot [mark=*, mark size=2] coordinates{(1.33,1.1)};
    \draw plot [mark=*, mark size=2] coordinates{(2.35,2.3)};
    
    \draw plot [mark=*, mark size=2] coordinates{(.8,0.65)}; 
    \draw plot [mark=*, mark size=2] coordinates{(-0.1,1.4)};
    
    \draw [-latex,dotted,thick] (2.35,2.3) --  node [right] {~} (-0.1,1.4); 
    \draw [-latex,dotted,thick] (-0.1,1.4) --  node [right] {~} (1.33,1.1); 
    \draw [-latex,dotted,thick] (1.33,1.1) --  node [right] {~} (.8,0.65); 
     \node[right, align=left, rotate=0]
at (1,0.45) {$x^*$};

     \node[right, align=left, rotate=0]
at (1.33,1.1) {$x_2$};

     \node[right, align=left, rotate=0]
at (2.35,2.3) {$x_1$};

\node[left, align=left, rotate=0]
at (.8,0.65) {$y_2$};
\node[left, align=left, rotate=0]
at (-0.1,1.4) {$y_1$};

\node[left, align=left, rotate=0]
at (0.3,0) {$B$};

\node[left, align=left, rotate=0]
at (2.7,.7) {$A$};

\end{tikzpicture}
    \caption{$A\cap B \neq \emptyset$.}
    \label{convexintersection}
    \end{subfigure}
    \begin{subfigure}[b]{0.5\textwidth}
\centering
\setlength\fheight{3cm}
\setlength\fwidth{7cm}
\begin{tikzpicture}

    \draw[rotate=45,fill=none] (0,0) ellipse (30pt and 65pt);

    \draw[rotate=-35,fill=none] (3.5,3) ellipse (30pt and 65pt);
    \draw plot [mark=*, mark size=2] coordinates{(3.35,0.45)}; 
    \draw plot [mark=*, mark size=2] coordinates{(3.83,1.2)};
    \draw plot [mark=*, mark size=2] coordinates{(4.9,2.2)};

    \draw plot [mark=*, mark size=2] coordinates{(1.33,0)};
    \draw plot [mark=*, mark size=2] coordinates{(.8,0.7)};
    \draw plot [mark=*, mark size=2] coordinates{(-0.2,1.5)};

    \draw [-latex,dotted,thick] (4.9,2.2) --  node [right] {~} (-0.2,1.5); 
    \draw [-latex,dotted,thick] (-0.2,1.5) --  node [right] {~} (3.83,1.2);
     \draw [-latex,dotted,thick] (3.83,1.2) --  node [right] {~} (.8,0.7);

    \node[right, align=left, rotate=0] at (3.35,0.45) {$x^*$};
    \node[right, align=left, rotate=0] at (3.83,1.2) {$x_2$};
    \node[right, align=left, rotate=0] at (4.9,2.1) {$x_1$};
    
     \node[left, align=left, rotate=0] at (1.33,0) {$y^*$};
     \node[left, align=left, rotate=0] at (.8,0.7) {$y_2$};
     \node[left, align=left, rotate=0] at (-0.2,1.4) {$y_1$};

\node[left, align=left, rotate=0] at (0.3,0) {$B$};
\node[left, align=left, rotate=0]at (5,.7) {$A$};

\end{tikzpicture}
    \caption{$A\cap B = \emptyset$.}
    \label{convexnointersection}
    \end{subfigure}
    \caption[map]{First few iterations of the Alternating Projections Method. (a) Both sequences are converging to the point $x^* \in A\cap B$. (b) Sequence $x_k$ is converging to $x^*\in A$, and the sequence $y_k$ is converging to $y^*\in B$, where $\|x^*-y^*\|_2 = \mathbf{dist(A,B)}$. }
\end{figure}
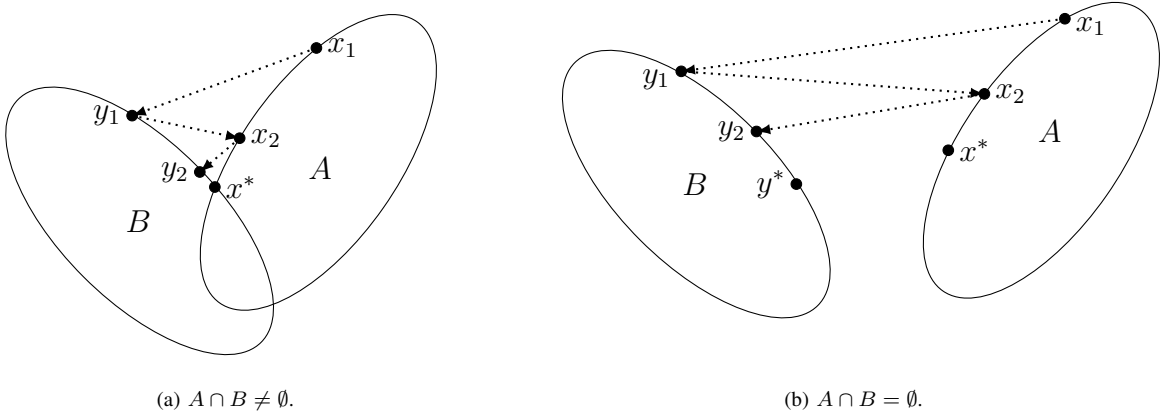

\textcolor{black}{ 1) $A\cap B \neq \emptyset$.
Alternating projections finds a point at the intersection of the sets, provided they intersect. We are not assuming the algorithm produces a point in $A\cap B$ in finite steps. We presume sequence $x_k \in A$, satisfies $\mathbf{dist}(x_k,B)\xrightarrow{}0$, and similarly for $y_k$.  See Fig.~\ref{convexintersection}.}

\textcolor{black}{ 2) $A\cap B = \emptyset$.
Alternating projections are also useful when the sets do not intersect. In this case, we can prove the following. Assume the distance between $A$ and $B$ is achieved (i.e., there exists points in $A$ and $B$ whose distance is $\mathbf{dist}(A,B)$). Then $x_k \xrightarrow{}x^*\in A$, $y_k \xrightarrow{}y^*\in B$, where $\|x^*-y^*\|_2 = \mathbf{dist}(A,B)$. In other words, alternating projections yields a pair of points in $A$ and $B$ that have minimum distance. In this case, alternating projections also yields (in the limit) a hyperplane that separates $A$ and $B$. See Fig. \ref{convexnointersection}.
\begin{proof}
The convergence proof of the alternating projections method is in Appendix B.
\end{proof}}


\subsection{Hybrid Analog/Digital Beamforming}
In this part, we decompose full-digital beamformers into their equivalent analog and digital components.
We add the CA constraints on the phase shifters.
The optimization problem is
\begin{equation}
\max_{\substack{\mathbf{W}_{\mathsf{BS}}^{\mathsf{RF}},\mathbf{F}_{\mathsf{BS}}^{\mathsf{RF}},\mathbf{F}_{\mathsf{UE}}^{\mathsf{RF}},\mathbf{W}_{\mathsf{UE}}^{\mathsf{RF}}\\\mathbf{W}_{\mathsf{BS}}^{\mathsf{BB}}[k],\mathbf{F}_{\mathsf{BS}}^{\mathsf{BB}}[k],\mathbf{F}_{\mathsf{UE}}^{\mathsf{BB}}[k],\mathbf{W}_{\mathsf{UE}}^{\mathsf{BB}}[k]\\k=0,\ldots,K-1}} \mathcal{I}(\rho_u) + \mathcal{I}(\rho_d) 
\end{equation}
\begin{equation*}
\begin{split}
\textsf{Subject to}~&\mathbf{W}_{\mathsf{BS}}[k] =\mathbf{W}_{\mathsf{BS}}^{\mathsf{RF}}\mathbf{W}_{\mathsf{BS}}^{\mathsf{BB}}[k] \\&
\mathbf{W}^*_{\mathsf{BS}}[k]\mathbf{W}_{\mathsf{BS}}[k] = \mathbf{I}_{N_{\mathsf{s}}} \\&
\mathbf{F}_{\mathsf{BS}}[k] =\mathbf{F}_{\mathsf{BS}}^{\mathsf{RF}}\mathbf{F}_{\mathsf{BS}}^{\mathsf{BB}}[k]
\\&
\mathbf{F}^*_{\mathsf{BS}}[k]\mathbf{F}_{\mathsf{BS}}[k] = \mathbf{I}_{N_{\mathsf{s}}} \\&
\mathbf{F}_{\mathsf{UE}}[k] =\mathbf{F}_{\mathsf{UE}}^{\mathsf{RF}}\mathbf{F}_{\mathsf{BS}}^{\mathsf{BB}}[k]
\\&
\mathbf{F}^*_{\mathsf{UE}}[k]\mathbf{F}_{\mathsf{UE}}[k] = \mathbf{I}_{N_{\mathsf{s}}} \\&
\mathbf{W}_{\mathsf{UE}}[k] =\mathbf{W}_{\mathsf{UE}}^{\mathsf{RF}}\mathbf{W}_{\mathsf{UE}}^{\mathsf{BB}}[k]
\\&
\mathbf{W}^*_{\mathsf{UE}}[k]\mathbf{W}_{\mathsf{UE}}[k] = \mathbf{I}_{N_{\mathsf{s}}} \\&
\mathbf{W}_{\mathsf{BS}}^{\mathsf{RF}*}\mathbf{G}_s\mathbf{F}_{\mathsf{BS}}^{\mathsf{RF}} = \mathbf{0}\\&
\mathbf{W}_{\mathsf{BS}}^{\mathsf{RF}},\mathbf{F}_{\mathsf{BS}}^{\mathsf{RF}}\in \mathbb{V}^{N_{\mathsf{BS}}\times N_{\mathsf{RF}}^{\mathsf{BS}}}\\&\mathbf{W}_{\mathsf{UE}}^{\mathsf{RF}},\mathbf{F}_{\mathsf{UE}}^{\mathsf{RF}}\in \mathbb{V}^{N_{\mathsf{UE}}\times N_{\mathsf{RF}}^{\mathsf{UE}}}\\&
k = 0,\ldots, K-1
\end{split}
\end{equation*}
where $\mathbf{G}_s$ is the SI subcarrier with lowest energy and $\mathbb{V}^{M\times L}$ is the set of feasible analog solutions that satisfy the CA constraints on the phase shifters. The ZF constraint is defined in the analog and not digital domain to cancel SI before downconversion, sampling and quantization to avoid ADC saturation due to SI. Like the full-digital case, the optimization problem has the form
\begin{equation}{\label{hybrid}}
    \max_{\mathbf{X}_\mathsf{BB},\mathbf{X}_\mathsf{RF}}~\log\det\left( \mathbf{I}_N + \rho\mathbf{X}_\mathsf{BB}^*\mathbf{X}^*_\mathsf{RF}\mathbf{A}\mathbf{A}^*\mathbf{X}_{\mathsf{RF}}\mathbf{X}_{\mathsf{BB}}\right)
\end{equation}
\begin{equation*}
\begin{split}
\textsf{Subject to}~&\mathbf{X}_\mathsf{BB}^*\mathbf{X}^*_\mathsf{RF}\mathbf{X}_{\mathsf{RF}}\mathbf{X}_{\mathsf{BB}} = \mathbf{I}_N\\&
\mathbf{X}_{\mathsf{RF}}^*\mathbf{C} = \mathbf{0}\\&
\mathbf{X}_{\mathsf{RF}} \in \mathbb{V}^{M\times L}
\end{split}
\end{equation*}

Note that $\mathbf{X}_{\mathsf{RF}} \in \mathbb{C}^{M\times L}$ and $\mathbf{X}_{\mathsf{BB}} \in \mathbb{C}^{L \times N}$ are the optimization variables, and $\mathbf{A} \in \mathbb{C}^{M \times N}$ and $\mathbf{C}\in \mathbb{C}^{M \times P}$ are given matrices. The analog beamformer should satisfy the ZF constraint $\mathbf{X}_{\mathsf{RF}}^*\mathbf{C} = \mathbf{0}$ {\em and} the CA constraint $\mathbf{X}_{\mathsf{RF}} \in \mathbb{V}^{M\times L}$. In fact, imperfect SI cancellation may introduce severe degradation in spectral efficiency. This might occur because the analog beamformers need to be projected onto the SI null subspace {\em and} CA subspace. The latter projection may violate the ZF constraint and hence the SI may not be perfectly eliminated. For example, authors in \cite{ca} discussed how the CA constraint violated the ZF condition and quantified losses in spectral efficiency. To circumvent this limitation, a common approach is to use \textit{Alternating Projections} explained \textcolor{black}{in Sec. \ref{altprojsec}}.  \textcolor{black}{This well-established method seeks a point at the intersection of two sets; when the sets are closed and convex, the method is known as projection onto convex sets and it converges to a point in the intersection \cite{alternatingprojections17}. In other cases, convergence is not guaranteed, but nevertheless
the method is routinely applied to nonconvex problems \cite{alternatingprojection18}.}

We alternately project the analog beamformers between the CA and ZF subspaces to find a common subspace that optimally meets both conditions. After convergence, the algorithm provides the optimal subspace that eliminates SI and maximizes sum spectral efficiency. To satisfy both conditions, we propose two nested iterative loops where the outer loop applies ZF cyclic maximization (like full-digital case) and the inner loop applies alternating projections. Because closed-form analog solutions do not exist for (\ref{hybrid}), the optimal digital solution $\mathbf{X}_{\mathsf{BB}}$ can be expressed in terms of the analog beamformers. Algorithm \ref{Hybrid-beamformingalgo} describes the hybrid beamforming design algorithm, and Table \ref{hybrid-beamforming-complexity} analyzes its computational complexity.
\textcolor{black}{\begin{proof}
Appendices B and C prove the analog and digital solutions, respectively.
\end{proof}}

\begin{algorithm}[H]
\caption{Hybrid Beamforming
\hfill \textcolor{blue}{Matrix Dimensions}
\label{Hybrid-beamformingalgo}}
\begin{algorithmic}[1]
\Function{Analog}{$\mathbf{A},\mathbf{C},L$} \funclabel{alg:a1}
    \hfill \textcolor{blue}{
            $\mathbf{A} \in\mathbb{C}^{M\times N};
            \mathbf{C} \in \mathbb{C}^{M\times L};
            M > N$}
\label{alg:a-line1}
    \State $\mathbf{X}_{\mathsf{RF}} \gets L~\text{Dominant left singular vectors \cite{svdComplexityAnalysis} of } \mathbf{A}$
        \hfill
        \textcolor{blue}{
            $ \mathbf{X}_{\mathsf{RF}} \in \mathbb{C}^{M\times L} $}
    \State $\mathbf{P}_{\bot} \gets \mathbf{I}-\mathbf{C}\mathbf{C}^{*}$
        \hfill
        \textcolor{blue}{
        $\mathbf{P}_{\bot} \in\mathbb{C}^{M\times M}$ }
    \For{$k\gets 1:N_{\mathsf{inner}}$}
    \State $\mathbf{Y}\gets\mathbf{P}_{\bot}\mathbf{X}_{\mathsf{RF}}$
        \hfill
        \textcolor{blue}{ $\mathbf{Y} \in \mathbb{C}^{M\times L}$ }
      \For{$i\gets1:M~\text{and}~j\gets1:L$}
         \State $\left(\mathbf{X}_{\mathsf{RF}}\right)_{ij} \gets \frac{ \left(\mathbf{Y} \right)_{ij}}{|\left(\mathbf{Y} \right)_{ij}|}$
      \EndFor
    \EndFor
    \State \Return $\mathbf{X}_{\mathsf{RF}}$
\EndFunction
\Statex
\Function{Digital}{$\mathbf{X}_{\mathsf{RF}},\mathbf{A},N$} \funclabel{alg:a2}
    \hfill \textcolor{blue}{$\mathbf{A}, \mathbf{X}_{\mathsf{RF}} \in\mathbb{C}^{M\times L}; M > L; M > N$}
\label{alg:a-line2}
    \State Compute SVD $\mathbf{X}_{\mathsf{RF}} = \mathbf{U}_{\mathsf{RF}} \mathbf{S}_{\mathsf{RF}}\mathbf{V}^*_{\mathsf{RF}}$
        \hfill \textcolor{blue}{$\mathbf{U}_{\mathsf{RF}} \in\mathbb{C}^{M\times M};
        \mathbf{S}_{\mathsf{RF}} \in \mathbb{C}^{M\times L};
        \mathbf{V}_{\mathsf{RF}} \in \mathbb{C}^{L\times L}$}
    \State $\mathbf{Q} \gets N~\text{Dominant left singular vectors of } \mathbf{U}^*_{\mathsf{RF}}\mathbf{A}$
        \hfill \textcolor{blue}{$\mathbf{Q} \in\mathbb{C}^{M\times N}$}
    \State $\mathbf{X}_{\mathsf{BB}} \gets \mathbf{V}_{\mathsf{RF}} \mathbf{S}^{-1}_{\mathsf{RF}} \mathbf{Q} $ 
            \hfill \textcolor{blue}{$\mathbf{X}_{\mathsf{BB}} \in\mathbb{C}^{N\times N}$}
    \State \Return $\mathbf{X}_{\mathsf{BB}}$
\EndFunction
\Statex
 \State \textbf{Input} $\mathbf{H}_s[k],\mathbf{H}_u[k],\mathbf{H}_d[k],~k=0,\ldots,K-1$ 
 \hfill
    \textcolor{blue}{
    $\mathbf{G}^{*}_d, \mathbf{G}_u, \mathbf{H}^{*}_d[k], \mathbf{H}_u[k] \in \mathbb{C}^{N_{\mathsf{BS}} \times N_{\mathsf{UE}}}$}
 \State Select the uplink/downlink subcarrier of highest energy
   \hfill
    \textcolor{blue}{
    $\mathbf{G}_s, \mathbf{H}_s[k] \in \mathbb{C}^{N_{\mathsf{BS}} \times N_{\mathsf{BS}}}$}
 \State $k^\star_u = \max\limits_{k = 0,\ldots,K-1}\|\mathbf{H}_u[k]\|_{F}^2$, $k^\star_d = \max\limits_{k = 0,\ldots,K-1}\|\mathbf{H}_d[k]\|_{F}^2$
 \State Select the SI subcarrier of lowest energy
 \State $k^\star_s = \min\limits_{k = 0,\ldots,K-1}\|\mathbf{H}_s[k]\|_{F}^2$
 \State $\mathbf{G}_u \gets \mathbf{H}_u[k^\star_u],\mathbf{G}_d \gets \mathbf{H}_d[k^\star_d]$  and $\mathbf{G}_s \gets \mathbf{H}_s[k^\star_s]$
\State \textbf{Initialize} $\mathbf{F}_{\mathsf{BS}}^{\mathsf{RF}},\mathbf{F}_{\mathsf{UE}}^{\mathsf{RF}},\mathbf{F}_{\mathsf{BS}}^{\mathsf{BB}}[k],\mathbf{F}_{\mathsf{UE}}^{\mathsf{BB}}[k],~k=0,\ldots,K-1$
    \hfill \textcolor{blue}{
    $ \mathbf{F}_{\mathsf{BS}}^{\mathsf{RF}} \in
      \mathbb{C}^{N_{\mathsf{BS}} \times N_{\mathsf{RF}}^{\mathsf{BS}}}; \mathbf{F}_{\mathsf{UE}}^{\mathsf{RF}} \in
      \mathbb{C}^{N_{\mathsf{UE}} \times N_{\mathsf{RF}}^{\mathsf{UE}}}$}
\State $\mathbf{F}_{\mathsf{UE}}[k] \gets  \mathbf{F}_{\mathsf{UE}}^{\mathsf{RF}}\mathbf{F}_{\mathsf{UE}}^{\mathsf{BB}}[k],~k=0,\ldots,K-1$
\hfill \textcolor{blue}{
    $ \mathbf{F}_{\mathsf{UE}}[k] \in 
      \mathbb{C}^{N_{\mathsf{UE}} \times N_{\mathsf{s}}}; 
      \mathbf{F}_{\mathsf{UE}}^{\mathsf{BB}}[k] \in
      \mathbb{C}^{N_{\mathsf{RF}}^{\mathsf{UE}} \times N_{\mathsf{s}}}$}
\State $\mathbf{F}_{\mathsf{BS}}[k] \gets  \mathbf{F}_{\mathsf{BS}}^{\mathsf{RF}}\mathbf{F}_{\mathsf{BS}}^{\mathsf{BB}}[k],~k=0,\ldots,K-1$ 
\hfill \textcolor{blue}{
    $ \mathbf{F}_{\mathsf{BS}}[k] \in 
      \mathbb{C}^{N_{\mathsf{BS}} \times N_{\mathsf{s}}}; 
      \mathbf{F}_{\mathsf{BS}}^{\mathsf{BB}}[k] \in
      \mathbb{C}^{N_{\mathsf{RF}}^{\mathsf{BS}} \times N_{\mathsf{s}}} $ }
\For{$t\gets 1:N_{\mathsf{outer}}$}
\State $\mathbf{W}_{\mathsf{BS}}^{\mathsf{RF}} \gets$ {\textsc{Analog}}$\left(\mathbf{G}_u\mathbf{F}_{\mathsf{UE}}[k^\star_u],\mathbf{G}_s\mathbf{F}_{\mathsf{BS}}^{\mathsf{RF}},N_{\mathsf{RF}}^{\mathsf{BS}}\right)$
    \hfill
    \textcolor{blue}{
    $ \mathbf{W}_{\mathsf{BS}}^{\mathsf{RF}} \in 
      \mathbb{C}^{N_{\mathsf{BS}} \times N_{\mathsf{RF}}^{\mathsf{BS}}} $ }
\State $\mathbf{W}_{\mathsf{BS}}^{\mathsf{BB}}[k] \gets$ {\textsc{Digital}}$\left(\mathbf{W}_{\mathsf{BS}}^{\mathsf{RF}},\mathbf{H}_u[k]\mathbf{F}_{\mathsf{UE}}[k],N_{\mathsf{s}}\right),~k=0,\ldots,K-1$
    \hfill
    \textcolor{blue}{
    $ \mathbf{W}_{\mathsf{BS}}^{\mathsf{BB}}[k] \in 
      \mathbb{C}^{N_{\mathsf{RF}}^{\mathsf{BS}} \times N_{\mathsf{s}}} $ }
\State $\mathbf{W}_{\mathsf{BS}}[k] \gets \mathbf{W}_{\mathsf{BS}}^{\mathsf{RF}}\mathbf{W}_{\mathsf{BS}}^{\mathsf{BB}}[k],~k=0,\ldots,K-1$
    \hfill
    \textcolor{blue}{
    $ \mathbf{W}_{\mathsf{BS}}[k] \in 
      \mathbb{C}^{N_{\mathsf{BS}} \times N_{\mathsf{s}}} $ }
\State $\mathbf{W}_{\mathsf{UE}}^{\mathsf{RF}}\gets N_{\mathsf{RF}}^{\mathsf{UE}}~\text{Dominant left singular vectors of } \mathbf{G}_d\mathbf{F}_{\mathsf{BS}}[k^\star_d]$ 
    \hfill
    \textcolor{blue}{
    $ \mathbf{W}_{\mathsf{UE}}^{\mathsf{RF}} \in 
      \mathbb{C}^{N_{\mathsf{UE}} \times N_{\mathsf{RF}}^{\mathsf{UE}}} $ }
\State $\mathbf{W}_{\mathsf{UE}}^{\mathsf{BB}}[k] \gets$ {\textsc{Digital}}$\left(\mathbf{W}_{\mathsf{UE}}^{\mathsf{RF}},\mathbf{H}_d[k]\mathbf{F}_{\mathsf{BS}}[k],N_{\mathsf{s}}\right),~k=0,\ldots,K-1$
    \hfill
    \textcolor{blue}{
    $ \mathbf{W}_{\mathsf{UE}}^{\mathsf{BB}}[k] \in
      \mathbb{C}^{N_{\mathsf{UE}} \times N_{\mathsf{s}}}$ }
\State $\mathbf{W}_{\mathsf{UE}}[k] \gets \mathbf{W}_{\mathsf{UE}}^{\mathsf{RF}}\mathbf{W}_{\mathsf{UE}}^{\mathsf{BB}}[k],~k=0,\ldots,K-1$
    \hfill
    \textcolor{blue}{
    $ \mathbf{W}_{\mathsf{UE}}[k] \in 
      \mathbb{C}^{N_{\mathsf{UE}} \times N_{\mathsf{s}}} $ }
\State $\mathbf{F}_{\mathsf{UE}}^{\mathsf{RF}}\gets N_{\mathsf{RF}}^{\mathsf{UE}}~\text{Dominant left singular vectors of } \mathbf{G}_u^*\mathbf{W}_{\mathsf{BS}}[k^\star_u]$ 
    \hfill
    \textcolor{blue}{
    $ \mathbf{F}_{\mathsf{UE}}^{\mathsf{RF}} \in 
      \mathbb{C}^{N_{\mathsf{UE}} \times N_{\mathsf{RF}}^{\mathsf{UE}}} $ }
\State $\mathbf{F}_{\mathsf{UE}}^{\mathsf{BB}}[k] \gets$ {\textsc{Digital}}$\left(\mathbf{F}_{\mathsf{UE}}^{\mathsf{RF}},\mathbf{H}^*_u[k]\mathbf{W}_{\mathsf{BS}}[k],N_{\mathsf{s}}\right),~k=0,\ldots,K-1$
    \hfill
    \textcolor{blue}{
    $ \mathbf{F}^{\mathsf{BB}}_{\mathsf{UE}}[k] \in 
      \mathbb{C}^{N_{\mathsf{RF}}^{\mathsf{UE}} \times N_{\mathsf{s}}} $ }
\State $\mathbf{F}_{\mathsf{UE}}[k] \gets \mathbf{F}_{\mathsf{UE}}^{\mathsf{RF}}\mathbf{F}_{\mathsf{UE}}^{\mathsf{BB}}[k],~k=0,\ldots,K-1$
    \hfill
    \textcolor{blue}{
    $ \mathbf{F}_{\mathsf{UE}}[k] \in 
      \mathbb{C}^{N_{\mathsf{UE}} \times N_{\mathsf{s}}} $ }
\State $\mathbf{F}_{\mathsf{BS}}^{\mathsf{RF}} \gets$ {\textsc{Analog}}$\left(\mathbf{G}_d^*\mathbf{W}_{\mathsf{UE}}[k^\star_d],\mathbf{G}^*_s\mathbf{W}_{\mathsf{BS}}^{\mathsf{RF}},N_{\mathsf{RF}}^{\mathsf{BS}}\right)$
    \hfill
    \textcolor{blue}{
    $ \mathbf{F}_{\mathsf{BS}}^{\mathsf{RF}} \in 
      \mathbb{C}^{N_{\mathsf{BS}} \times N_{\mathsf{RF}}^{\mathsf{BS}}} $ }
\State $\mathbf{F}_{\mathsf{BS}}^{\mathsf{BB}}[k] \gets$ {\textsc{Digital}}$\left(\mathbf{F}_{\mathsf{BS}}^{\mathsf{RF}},\mathbf{H}^*_d[k]\mathbf{W}_{\mathsf{UE}}[k],N_{\mathsf{s}}\right),~k=0,\ldots,K-1$
    \hfill
    \textcolor{blue}{
    $ \mathbf{F}_{\mathsf{BS}}^{\mathsf{BB}}[k] \in 
      \mathbb{C}^{N_{\mathsf{RF}} \times N_{\mathsf{s}}} $ }
\State $\mathbf{F}_{\mathsf{BS}}[k] \gets \mathbf{F}_{\mathsf{BS}}^{\mathsf{RF}}\mathbf{F}_{\mathsf{BS}}^{\mathsf{BB}}[k],~k=0,\ldots,K-1$
    \hfill
    \textcolor{blue}{
    $ \mathbf{F}_{\mathsf{BS}}[k] \in 
      \mathbb{C}^{N_{\mathsf{BS}} \times N_{\mathsf{s}}} $ }
\EndFor
\State \Return $\mathbf{F}_{\mathsf{BS}}[k],\mathbf{W}_{\mathsf{BS}}[k],\mathbf{W}_{\mathsf{UE}}[k],\mathbf{F}_{\mathsf{UE}}[k],~k=0,\ldots,K-1$
\end{algorithmic}
\end{algorithm}

\begingroup
\renewcommand\arraystretch{0.6}
  \centering 
\textcolor{black}{\begin{longtable}{cc}
\caption{Computational Complexity of the Hybrid Beamforming Algorithm.
\label{hybrid-beamforming-complexity}}\\
    \bfseries Operation & \bfseries Complex Multiplications for Highest-Order Terms \\
    \hline
    \multicolumn{2}{c}{\textsc{Analog}($\mathbf{A},\mathbf{C},L$)
    \hfill
    \textcolor{black}{
            $\mathbf{A} \in\mathbb{C}^{M\times N};
            \mathbf{C}, \mathbf{X}_{\mathsf{RF}}, \mathbf{Y} \in \mathbb{C}^{M\times L};
            \mathbf{P}_{\bot} \in\mathbb{C}^{M\times M};
            M > N$}}\\
    \hline
    $\mathbf{X}_{\mathsf{RF}}$ & $9 M N^2$\\
    $\mathbf{P}_{\bot}$ & $L M^2$\\
    Inner Loop & $N_{\mathsf{inner}} L M^2$ \\
    \bfseries Overall & 
    $9 M N^2+ \left( N_{\mathsf{inner}} + 1 \right) L M^2$\\
    \hline
    \multicolumn{2}{c}{\textsc{Digital}($\mathbf{X}_{\mathsf{RF}},\mathbf{A},N$)
    \hfill
    $\mathbf{A}, \mathbf{X}_{\mathsf{RF}}  \in\mathbb{C}^{M\times L}; \mathbf{Q} \in\mathbb{C}^{M\times N}; 
    \mathbf{X}_{\mathsf{BB}} \in\mathbb{C}^{N\times N};
    M > L; M > N$}\\
    \hline
    SVD of $\mathbf{X}_{\mathsf{RF}}$ & $9 M L^2$\\
    $\mathbf{Q}$ & $N M^2 + 9 M N^2$\\
    $\mathbf{X}_{\mathsf{BB}}$ & $N^3$\\
    \bfseries Overall & $9 M L^2 + 9 M N^2 + N M^2 + N^3$\\
    \hline
    \multicolumn{2}{c}{\textsc{Algorithm}}\\
    \hline
    Max, Min indices & 
        $K N_{\mathsf{BS}}^2 + 2 K N_{\mathsf{BS}} N_{\mathsf{UE}}$\\
    Set $\left[\mathbf{F}_{\mathsf{UE}}[k]\right]_{k=0}^{K-1}$ &
        $K N_{\mathsf{RF}}^{\mathsf{UE}} N_{\mathsf{s}} N_{\mathsf{UE}}$\\
    Set $\left[\mathbf{F}_{\mathsf{BS}}[k]\right]_{k=0}^{K-1}$ &
        $K N_{\mathsf{BS}} N_{\mathsf{RF}}^{\mathsf{BS}} N_{\mathsf{s}}$\\
    Outer Loop, Each Iteration: \\
        $\mathbf{G}_u\mathbf{F}_{\mathsf{UE}}[k^\star_u]$ & $N_{\mathsf{BS}} N_{\mathsf{s}} N_{\mathsf{UE}}$\\
    $\mathbf{G}_s\mathbf{F}_{\mathsf{BS}}^{\mathsf{RF}}$ &
        $ N_{\mathsf{RF}}^{\mathsf{BS}} N_{\mathsf{BS}}^2 $\\
    $\mathbf{W}^{\mathsf{RF}}_{\mathsf{BS}}$ &  
        $9 N_{\mathsf{BS}} N_\mathsf{s}^2 +
         \left( N_{\mathsf{inner}} + 1 \right) N_{\mathsf{RF}}^{\mathsf{BS}} N_{\mathsf{BS}}^2  $ \\
    $\mathbf{H}_u[k]\mathbf{F}_{\mathsf{UE}}[k]$ &            
        $N_{\mathsf{BS}} N_\mathsf{s} N_{\mathsf{UE}}$ \\
    $\left[\mathbf{W}_{\mathsf{BS}}^{\mathsf{BB}}[k]\right]_{k=0}^{K-1}$ &
        $9 K N_{\mathsf{BS}} (N_{\mathsf{RF}}^{\mathsf{BS}})^2 + 
         9 K N_{\mathsf{BS}} N_\mathsf{s}^2 + K N_\mathsf{s} N_{\mathsf{BS}}^2 +
         K N_\mathsf{s}^3$ \\
    $\left[\mathbf{W}_{\mathsf{BS}}[k]\right]_{k=0}^{K-1}$ &
        $K N_{\mathsf{BS}} N_{\mathsf{RF}}^{\mathsf{BS}} N_\mathsf{s}$ \\
    $\mathbf{G}_d\mathbf{F}_{\mathsf{BS}}[k^\star_d]$ & 
        $N_{\mathsf{BS}} N_\mathsf{s} N_{\mathsf{UE}}$\\
    $\mathbf{W}_{\mathsf{UE}}^{\mathsf{RF}}$ & 
        $N_{\mathsf{BS}} N_\mathsf{s} N_{\mathsf{UE}} +
         9 N_{\mathsf{UE}} N_\mathsf{s}^2$  \\
    $\mathbf{H}_d[k]\mathbf{F}_{\mathsf{BS}}[k]$ & 
        $N_{\mathsf{BS}} N_\mathsf{s} N_{\mathsf{UE}}$\\
    $\left[\mathbf{W}_{\mathsf{UE}}^{\mathsf{BB}}[k]\right]_{k=0}^{K-1}$ & 
        $9 K N_{\mathsf{UE}} (N_{\mathsf{RF}}^{\mathsf{UE}})^2 + 9 K N_{\mathsf{UE}} N_\mathsf{s}^2 + K N_\mathsf{s} N_{\mathsf{UE}}^2 + K N_\mathsf{s}^3$ \\
    $\left[\mathbf{W}_{\mathsf{UE}}[k]\right]_{k=0}^{K-1}$ & 
        $K N_{\mathsf{RF}}^{\mathsf{UE}} N_\mathsf{s} N_{\mathsf{UE}}$\\
    $\mathbf{G}_u^*\mathbf{W}_{\mathsf{BS}}[k^\star_u]$& 
        $N_{\mathsf{BS}} N_\mathsf{s} N_{\mathsf{UE}}$\\
    $\mathbf{F}_{\mathsf{UE}}^{\mathsf{RF}}$ & 
        $N_{\mathsf{BS}} N_\mathsf{s} N_{\mathsf{UE}} + 
         9 N_{\mathsf{UE}} N_\mathsf{s}^2$ \\
    $\mathbf{H}^*_u[k]\mathbf{W}_{\mathsf{BS}}[k]$ & 
        $N_{\mathsf{BS}} N_\mathsf{s} N_{\mathsf{UE}}$\\
    $\left[\mathbf{F}_{\mathsf{UE}}^{\mathsf{BB}}[k]\right]_{k=0}^{K-1}$ &
        $9 K N_{\mathsf{UE}} (N_{\mathsf{RF}}^{\mathsf{UE}})^2 + 9 K N_{\mathsf{UE}} N_\mathsf{s}^2 + K N_\mathsf{s} N_{\mathsf{UE}}^2 + K N_\mathsf{s}^3$ \\
    $\left[\mathbf{F}_{\mathsf{UE}}[k]\right]_{k=0}^{K-1}$ & 
        $K N_{\mathsf{UE}} N_{\mathsf{RF}}^{\mathsf{UE}} N_\mathsf{s}$\\
    $\mathbf{G}_d^*\mathbf{W}_{\mathsf{UE}}[k^\star_d]$ & 
        $N_{\mathsf{BS}} N_\mathsf{s} N_{\mathsf{UE}}$\\
    $\mathbf{G}_s^*\mathbf{W}_{\mathsf{BS}}^{\mathsf{RF}}$ &         
        $N_{\mathsf{RF}}^{\mathsf{BS}} N_{\mathsf{BS}}^2$\\
    $\mathbf{F}_{\mathsf{BS}}^{\mathsf{RF}}$ & 
        $9 N_{\mathsf{BS}} N_{\mathsf{s}}^2+ \left( N_{\mathsf{inner}} + 1 \right) N_{\mathsf{RF}}^{\mathsf{BS}} N_{\mathsf{BS}}^2$ \\
    $\mathbf{H}_d^*[k]\mathbf{W}_{\mathsf{UE}}[k]$ & 
        $N_{\mathsf{BS}} N_\mathsf{s} N_{\mathsf{UE}}$\\
    $\left[\mathbf{F}_{\mathsf{BS}}^{\mathsf{BB}}[k]\right]_{k=0}^{K-1}$ &
        $9 K N_{\mathsf{BS}} (N_{\mathsf{RF}}^{\mathsf{BS}})^2 + 
         9 K N_{\mathsf{BS}} N_\mathsf{s}^2 + 
         K N_\mathsf{s} N_{\mathsf{BS}}^2 + K N_\mathsf{s}^3$ \\
    $\left[\mathbf{F}_{\mathsf{BS}}[k]\right]_{k=0}^{K-1}$ &
        $K N_{\mathsf{BS}} N_{\mathsf{RF}}^{\mathsf{BS}} N_\mathsf{s}$
    \\
    \hline
\end{longtable}}
\endgroup


From the computational complexity analysis of the highest-order terms in Algorithm \ref{Hybrid-beamformingalgo} given in Table \ref{hybrid-beamforming-complexity}, the overall number of complex multiplications is 
\begin{equation*}
\begin{split}
    9 N_{\mathsf{outer}}
      N_{\mathsf{BS}} N_{\mathsf{s}} N_{\mathsf{UE}} +
    K \left( N_{\mathsf{outer}} + 1 \right)
      N_{\mathsf{RF}}^\mathsf{UE} N_{\mathsf{s}} N_{\mathsf{UE}} + 
    K N_{\mathsf{BS}}^2 + 2 K N_{\mathsf{BS}} + \\
    \left( 2 N_{\mathsf{inner}} + 4 \right) N_{\mathsf{outer}}
      N_{\mathsf{RF}}^{\mathsf{BS}} N_{\mathsf{BS}}^2 +
    18 N_{\mathsf{outer}}
      N_{\mathsf{BS}} N_{\mathsf{s}}^2 +
    K \left( 2 N_{\mathsf{outer}} + 1 \right)
      N_{\mathsf{BS}} N_{\mathsf{RF}}^{\mathsf{BS}} N_{\mathsf{s}} + \\
    28 N_{\mathsf{outer}} 
      N_{\mathsf{UE}} N_{\mathsf{s}}^2 +
    2 K N_{\mathsf{outer}} 
      N_{\mathsf{s}} N_{\mathsf{UE}}^2 + 
    2 K N_{\mathsf{outer}} 
      N_{\mathsf{s}}^3 +
    18 K N_{\mathsf{outer}} 
      N_{\mathsf{UE}} \left( N_{\mathsf{RF}}^{\mathsf{UE}} \right)^2 + \\
    2 K N_{\mathsf{outer}}
      N_{\mathsf{s}}^3 +
    2 K N_{\mathsf{outer}} 
      N_{\mathsf{s}} N_{\mathsf{BS}}^2 +
    18 K N_{\mathsf{outer}} 
      N_{\mathsf{BS}} N_{\mathsf{s}}^2 +
    18 K N_{\mathsf{outer}} 
      N_{\mathsf{BS}} \left( N_{\mathsf{RF}}^{\mathsf{BS}} \right)^2
\end{split}    
\end{equation*}
When using the largest parameter values in Table \ref{param} of
$K=32$,
$N_{\mathsf{BS}}=16$,
$N_{\mathsf{inner}}=20$,
$N_{\mathsf{outer}}=10$, 
$N_{\mathsf{RF}}^{\mathsf{BS}}=2$,
$N_{\mathsf{RF}}^{\mathsf{UE}}=2$, 
$N_{\mathsf{s}}=2$, and
$N_{\mathsf{UE}}=4$, 
1.82M complex multiplications would be required.
88\% of the multiplications come from the following terms in descending order:
$2 K N_{\mathsf{outer}} 
  N_{\mathsf{s}} N_{\mathsf{BS}}^2 +
18 K N_{\mathsf{outer}} 
  N_{\mathsf{BS}} N_{\mathsf{s}}^2 +
18 K N_{\mathsf{outer}} 
  N_{\mathsf{BS}} \left( N_{\mathsf{RF}}^{\mathsf{BS}} \right)^2 +
\left( 2 N_{\mathsf{inner}} + 4 \right) N_{\mathsf{outer}}
  N_{\mathsf{RF}}^{\mathsf{BS}} N_{\mathsf{BS}}^2$.

\section{Performance Analysis}
\subsection{Energy Efficiency}
The energy efficiency, expressed in bits/s/Hz/Watt or bits/J/Hz, is defined as the ratio between the spectral efficiency and total power consumption. It is expressed as
\begin{equation}
\mathcal{J}(\rho) = \frac{\mathcal{I}(\rho)}{\rho_{\mathsf{Total}}}
\end{equation}
where $\rho_{\mathsf{Total}}$ is the total power consumption for full-digital (DC) and hybrid combiners (HC)
\begin{equation}
\rho^{\mathsf{DC}}_{\mathsf{Total}} =   N_{\mathsf{RX}} \left(\rho_{\mathsf{LNA}} + \rho_{\mathsf{RF}} + 2\rho_{\mathsf{ADC}}  \right) 
\end{equation}
\begin{equation}
\rho^{\mathsf{HC}}_{\mathsf{Total}} = N_{\mathsf{RX}}\left( \rho_{\mathsf{LNA}} + \rho_{\mathsf{SP}} + N_{\mathsf{RF}}\rho_{\mathsf{PS}}  \right) +  N_{\mathsf{RF}} \left( \rho_{\mathsf{RF}} + \rho_{\mathsf{C}} + 2\rho_{\mathsf{ADC}} \right)   
\end{equation}
Here, $\rho_{\mathsf{RF}}$ is the power consumption per RF chain which is defined by
$\rho_{\mathsf{RF}} = \rho_{\mathsf{M}} + \rho_{\mathsf{LO}} + \rho_{\mathsf{LPF}} + \rho_{\mathsf{B}\mathsf{B}_{\mathsf{amp}} }$.
Please see Table \ref{power} to example power consumption values for each subsystem.
\begin{table}[h]
\renewcommand{\arraystretch}{1}
\caption{Power Consumption of Each Subsystem \cite{adc}.}
\label{power}
\centering
\begin{tabular}{rll}
\bfseries Device & \bfseries Notation & \bfseries Value\\
\hline
Low Noise Amplifier (LNA) \cite{adc36} & $\rho_{\mathsf{LNA}}$ & 39 mW\\
Splitter & $\rho_{\mathsf{SP}}$ & 19.5 mW\\
Combiner \cite{adc36} & $\rho_{\mathsf{C}}$ & 19.5 mW\\
Phase Shifter \cite{adc37,adc38} & $\rho_{\mathsf{PS}}$ & 2 mW\\
Mixer \cite{adc39} & $\rho_{\mathsf{M}}$  & 16.8 mW\\
Local Oscillator \cite{adc27} & $\rho_{\mathsf{LO}}$ & 5 mW\\
Low Pass Filter \cite{adc27} & $\rho_{\mathsf{LPF}}$ & 14 mW\\
Baseband Amplifier \cite{adc27} & $\rho_{\mathsf{B}\mathsf{B}_{\mathsf{amp}} }$ & 5 mW\\
ADC & $\rho_{\mathsf{ADC}}$ & Table \ref{setting}
\end{tabular}
\end{table}
\subsection{Outage Probability}
Given a transmission strategy, the outage probability for rate $R$ (bits/s/Hz) is then
\begin{equation}
    P_{\textsf{out}}(\textsf{\scriptsize{SNR}}, R) = \mathbb{P}[\mathcal{I}(\textsf{\scriptsize{SNR}})<R].
\end{equation}
With powerful channel codes, probability of error when there is no outage is very small; hence, outage probability accurately approximates the actual block error probability. Modern radio systems such as UMTS and LTE operate at a target error probability. The primary performance measure is the maximum rate\footnote[2]{In this work, we define rate with outage as the average data rate correctly decoded at the receiver which is equivalent to throughput. In other approaches, the rate with outage is assimilated into the transmit data rate. For our approach, we account for the probability of bursts (outage) and multiply by ($1-\epsilon$); for the transmit data rate approach, the term ($1-\epsilon$) is not included.} at each {\textsf{\scriptsize{SNR}}}, such that the outage probability is less than $\epsilon$:
\begin{equation}
R_\epsilon(\textsf{\scriptsize{SNR}}) = \max_{\zeta}\left\{ \zeta: P_{\textsf{out}}(\textsf{\scriptsize{SNR}}, \zeta) \leq \epsilon \right\}
\end{equation}

\subsection{Upper Bound on Spectral Efficiency}
For interference-free scenarios, optimal beamformers diagonalize the channel. By applying SVD on each subcarrier matrix, the singular values are in descending order and we extract the first $N_{\mathsf{s}}$ singular vectors associated with the spatial streams. Equivalently, the upper bound is
\begin{equation}
\mathcal{I}({\scriptsize{\textsf{SNR}}}) =  \frac{1}{K}\sum_{k=0}^{K-1}\sum_{\ell=0}^{N_{\mathsf{s}}-1}\log\left(1 + \frac{{\scriptsize{\textsf{SNR}}}}{KN_{\mathsf{s}}} \, \sigma_{\ell}\left(\mathbf{H}[k] \right)^2  \right)  
\end{equation}

\textcolor{black}{
\subsection{Spectral Efficiency Gain and Loss}
We evaluate average achievable gain $\Gamma$ in FD rate $\mathcal{I}_{\mathsf{FD}}$ relative to rate $\mathcal{I}_{\mathsf{X}}$ by method $X$ as
\begin{equation}
\Gamma[\%] = \left(\frac{\mathcal{I}_{\mathsf{FD}}}{\mathcal{I}_{\mathsf{X}}} - 1  \right) \times 100.    
\end{equation}}
\textcolor{black}{
and the average achievable loss $\zeta$ in FD rate $\mathcal{I}_{\mathsf{FD}}$ relative to upper bound $\mathcal{I}_{\mathsf{Bound}}$ as
\begin{equation}
\zeta[\%] = \left(\frac{\mathcal{I}_{\mathsf{Bound}}}{\mathcal{I}_{\mathsf{FD}}} - 1  \right) \times 100.    
\end{equation}}

\section{Numerical Results}
In this section, we discuss numerical results of communication performance measures obtained by Monte Carlo simulation of 1000 realizations in each case in MATLAB. Unless otherwise stated, simulations used the parameter values in Table \ref{param}. 
\begin{table}[h]
\renewcommand{\arraystretch}{1}
\caption{System Parameters \cite{anum}.}
\label{param}
\centering
\begin{tabular}{rll}
\bfseries Parameter & \bfseries mmWave & \bfseries Sub-6 GHz\\
\hline
Carrier frequency & 28 GHz & 3.5 GHz\\
Bandwidth &850 MHz& 150 MHz\\
Number UE antennas ($N_{\mathsf{UE}}$) & 4& 2\\
Number relay/BS antennas ($N_{\mathsf{BS}}$) & 16TX+16RX & 8TX+8RX \\
Antenna separation & $\lambda/2$&$\lambda/2$\\
Antenna correlation & None & None\\
Number clusters ($C$) & 4 & 10\\
Number rays per cluster $(R_c)$ & 10 & 20\\
Angular spread& 2$^{\circ}$ & 2$^{\circ}$\\
Pathloss exponent ($\gamma$) & 3&3\\
Number of subcarriers ($K$) & 16& 32\\
Cyclic prefix length ($L_c$) & $K/4$ & $K/4$\\
Number BS RF chains ($N_{\mathsf{RF}}^{\mathsf{BS}}$) & 2 & 2\\
Number UE RF chains ($N_{\mathsf{RF}}^{\mathsf{UE}}$) & 2 & 2\\
Number spatial streams ($N_{\mathsf{s}}$) & 2 & 2\\
Angle between FD arrays & $\pi/2$ &  $\pi/2$\\
Distance between FD arrays & $2\lambda$ & $2\lambda$\\
Signal-to-Interference Ratio & -120 dB& -120 dB\\
Rician factor ($\kappa$) & 5 dB & 5 dB\\
Raised cosine roll-off factor & 1 & 1\\
Outer iterations ($N_{\text{outer}}$) & 10 & 10
\end{tabular}
\end{table}

\subsection{Primary Evaluation }
\begin{figure}[tb]
\centering
\setlength\fheight{5.5cm}
\setlength\fwidth{7.5cm}
\input{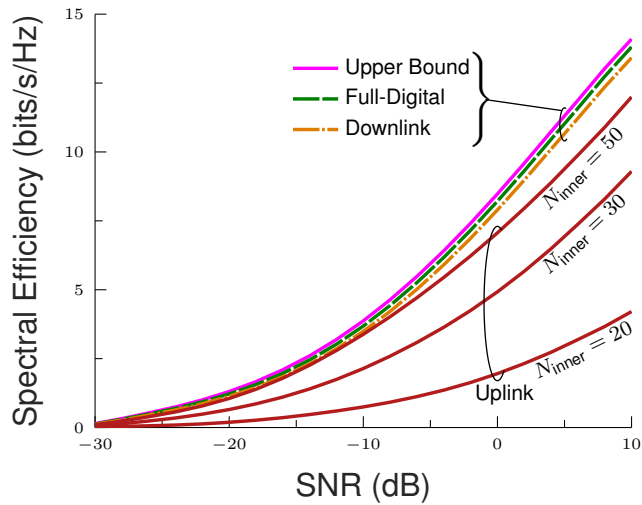}
    \caption{Spectral efficiency results: Performance is evaluated in terms of uplink and downlink spectral efficiency for hybrid architecture. The uplink rate is evaluated for different number of iterations of the inner loop corresponding to alternating projections. In addition, full-digital and upper bound are evaluated and serve as benchmarking tools. MmWave configuration is considered for this simulation.}
    \label{pict3}
\end{figure}

Fig. \ref{pict3} plots spectral efficiency vs.\ average SNR. We observe the full-digital beamformer designed using $N_{\text{outer}}=10$ iterations is robust to SI and very close to the upper bound.  In the downlink, which is interference-free, the hybrid beamformer is very close in performance for the full-digital beamformer. The small gap is due to phase shifter CA constraints. In the uplink, unlike the full-digital case, the hybrid beamformer shows some limitations; the performance improves with increasing inner alternating projections. The low performance at $N_{\text{inner}}=10$ iterations is due to the zero-forcing constraint being approximately but not completely satisfied; hence, the optimal precoders and combiners are not yet obtained. At $N_{\text{inner}}=50$ iterations, the beamformers are closely projected onto the ZF null-space, which further enhances the uplink rate. The results are consistent with those for machine-to-machine narrowband FD systems \cite{valcarce}.

\subsection{Alternating Projections and Signal-to-Interference Ratio}

\begin{figure}[tb]
\centering
\setlength\fheight{5.5cm}
\setlength\fwidth{7.5cm}
\input{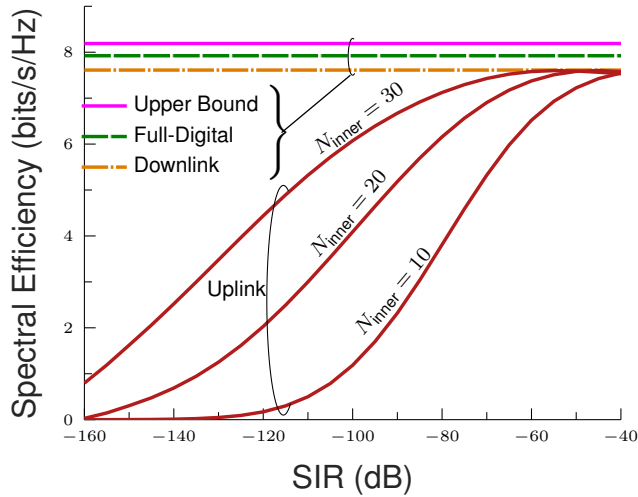}
    \caption{Spectral efficiency results: Performance at SNR = 0 dB and across a range of SIR from -160 to -40 dB which corresponds to cell-edge user (-160 dB), middle user (around -80 dB) and near user (-40 dB), respectively. Algorithm robustness is tested by varying the number of iterations of alternating projections. Note that mmWave setting is assumed for this scenario.}
    \label{pict5}
\end{figure}

Fig. \ref{pict5} presents spectral efficiency for different numbers of inner loop alternating projections across a range of SIRs. The full-digital beamformer remains close to the upper bound while the downlink rate for the hybrid architecture is still limited by CA phase shifter constraints. In accordance with conclusions drawn from Fig. \ref{pict3}, uplink rate improves with increasing inner loop alternating projections. We further observe the design is sensitive to high SI power (SIR from -160 to -140 dB) when using a few inner iterations. However, the performance can be improved by using more alternating projections. As SIR increases to around -40 dB, which corresponds to a user near the BS, the uplink performance matches the downlink (interference-free) performance regardless of the number of iterations.

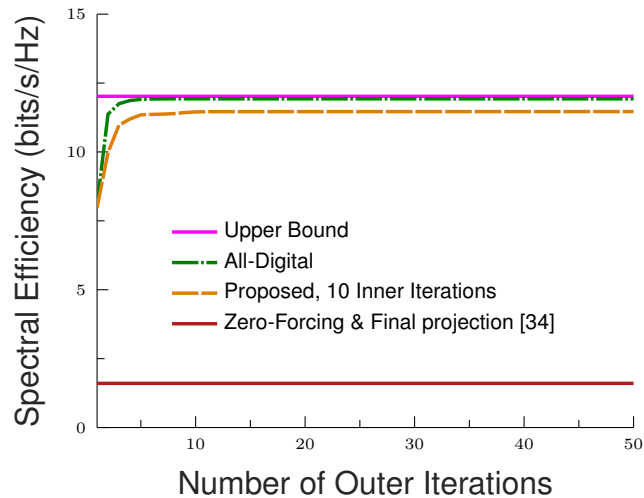
\begin{figure}[tb]
\centering
\setlength\fheight{5.5cm}
\setlength\fwidth{7.5cm}
%
%

\definecolor{mycolor1}{rgb}{1.00000,0.00000,1.00000}%
\definecolor{mycolor2}{rgb}{0.00000,0.49804,0.00000}%
\definecolor{mycolor3}{rgb}{0.00000,0.44706,0.74118}%
\definecolor{mycolor4}{rgb}{0.87059,0.49020,0.00000}%
\definecolor{cornellred}{rgb}{0.7, 0.11, 0.11}
\begin{tikzpicture}

\begin{axis}[%
width=0.951\fwidth,
height=\fheight,
at={(0\fwidth,0\fheight)},
scale only axis,
xmin=1,
xmax=50,
xlabel style={font=\color{white!15!black}},
xlabel={\sffamily{Number of Outer Iterations}},
ymin=0,
ymax=15,
ylabel style={font=\color{white!15!black}},
ylabel={\sffamily{Spectral Efficiency (bits/s/Hz) }},
axis x line*=bottom,
axis y line*=left,
axis background/.style={fill=white},
legend style={at={(0.5,0.2)}, anchor=south, legend cell align=left, align=left, draw=none}
]
\addplot [color=mycolor1, line width=1.3pt]
  table[row sep=crcr]{%
1	12.0193543937599\\
2	12.0193543937599\\
3	12.0193543937599\\
4	12.0193543937599\\
5	12.0193543937599\\
6	12.0193543937599\\
7	12.0193543937599\\
8	12.0193543937599\\
9	12.0193543937599\\
10	12.0193543937599\\
11	12.0193543937599\\
12	12.0193543937599\\
13	12.0193543937599\\
14	12.0193543937599\\
15	12.0193543937599\\
16	12.0193543937599\\
17	12.0193543937599\\
18	12.0193543937599\\
19	12.0193543937599\\
20	12.0193543937599\\
21	12.0193543937599\\
22	12.0193543937599\\
23	12.0193543937599\\
24	12.0193543937599\\
25	12.0193543937599\\
26	12.0193543937599\\
27	12.0193543937599\\
28	12.0193543937599\\
29	12.0193543937599\\
30	12.0193543937599\\
31	12.0193543937599\\
32	12.0193543937599\\
33	12.0193543937599\\
34	12.0193543937599\\
35	12.0193543937599\\
36	12.0193543937599\\
37	12.0193543937599\\
38	12.0193543937599\\
39	12.0193543937599\\
40	12.0193543937599\\
41	12.0193543937599\\
42	12.0193543937599\\
43	12.0193543937599\\
44	12.0193543937599\\
45	12.0193543937599\\
46	12.0193543937599\\
47	12.0193543937599\\
48	12.0193543937599\\
49	12.0193543937599\\
50	12.0193543937599\\
};
\addlegendentry{\sffamily{Upper Bound}}

\addplot [color=mycolor2, dash pattern={on 10pt off 1pt on 1pt off 1pt} , line width=1.3pt]
  table[row sep=crcr]{%
1	8.08216434246347\\
2	11.3719519486009\\
3	11.7492413926849\\
4	11.8663418268893\\
5	11.9071573388221\\
6	11.9204067583532\\
7	11.924659789642\\
8	11.9260844421295\\
9	11.9265906569975\\
10	11.926780118448\\
11	11.9268537483699\\
12	11.926883077452\\
13	11.9268949405098\\
14	11.9268997834894\\
15	11.9269017714863\\
16	11.9269025901806\\
17	11.9269029279693\\
18	11.9269030674902\\
19	11.9269031251538\\
20	11.9269031489942\\
21	11.9269031588526\\
22	11.9269031629296\\
23	11.9269031646158\\
24	11.9269031653131\\
25	11.9269031656015\\
26	11.9269031657208\\
27	11.9269031657702\\
28	11.9269031657906\\
29	11.926903165799\\
30	11.9269031658025\\
31	11.9269031658039\\
32	11.9269031658045\\
33	11.9269031658048\\
34	11.9269031658049\\
35	11.9269031658049\\
36	11.9269031658049\\
37	11.9269031658049\\
38	11.9269031658049\\
39	11.9269031658049\\
40	11.9269031658049\\
41	11.9269031658049\\
42	11.9269031658049\\
43	11.9269031658049\\
44	11.9269031658049\\
45	11.9269031658049\\
46	11.9269031658049\\
47	11.9269031658049\\
48	11.9269031658049\\
49	11.9269031658049\\
50	11.9269031658049\\
};
\addlegendentry{\sffamily{All-Digital}}


\addplot [color=mycolor4, dash pattern={on 10pt off 1pt on 0pt off 0pt} , line width=1.3pt]
  table[row sep=crcr]{%
1	7.97801902323639\\
2	9.97851160563203\\
3	10.9709646541203\\
4	11.1936300431128\\
5	11.346102816754\\
6	11.3628545805797\\
7	11.3732302292811\\
8	11.3904547457356\\
9	11.428770503789\\
10	11.4528986467056\\
11	11.4598919090427\\
12	11.4618216886926\\
13	11.4624430026336\\
14	11.4626876465042\\
15	11.4628092667073\\
16	11.4628878284436\\
17	11.4629523345141\\
18	11.4630143372815\\
19	11.4630785091081\\
20	11.4631465315106\\
21	11.4632187292114\\
22	11.4632948216878\\
23	11.4633742856479\\
24	11.4634565317833\\
25	11.4635409887192\\
26	11.4636271396098\\
27	11.463714534563\\
28	11.4638027910171\\
29	11.4638915884883\\
30	11.4639806610855\\
31	11.464069789558\\
32	11.4641587937536\\
33	11.4642475258796\\
34	11.4643358646938\\
35	11.4644237106166\\
36	11.4645109816876\\
37	11.4645976102603\\
38	11.4646835403258\\
39	11.4647687253579\\
40	11.4648531265865\\
41	11.4649367116161\\
42	11.4650194533223\\
43	11.4651013289676\\
44	11.465182319492\\
45	11.4652624089407\\
46	11.4653415840001\\
47	11.4654198336184\\
48	11.4654971486931\\
49	11.4655735218119\\
50	11.4656489470348\\
};
\addlegendentry{\sffamily{Proposed, 10 Inner Iterations}}

\addplot [color=cornellred, line width=1.3pt]
  table[row sep=crcr]{%
1	1.6026447203128\\
2	1.6026447203128\\
3	1.6026447203128\\
4	1.6026447203128\\
5	1.6026447203128\\
6	1.6026447203128\\
7	1.6026447203128\\
8	1.6026447203128\\
9	1.6026447203128\\
10	1.6026447203128\\
11	1.6026447203128\\
12	1.6026447203128\\
13	1.6026447203128\\
14	1.6026447203128\\
15	1.6026447203128\\
16	1.6026447203128\\
17	1.6026447203128\\
18	1.6026447203128\\
19	1.6026447203128\\
20	1.6026447203128\\
21	1.6026447203128\\
22	1.6026447203128\\
23	1.6026447203128\\
24	1.6026447203128\\
25	1.6026447203128\\
26	1.6026447203128\\
27	1.6026447203128\\
28	1.6026447203128\\
29	1.6026447203128\\
30	1.6026447203128\\
31	1.6026447203128\\
32	1.6026447203128\\
33	1.6026447203128\\
34	1.6026447203128\\
35	1.6026447203128\\
36	1.6026447203128\\
37	1.6026447203128\\
38	1.6026447203128\\
39	1.6026447203128\\
40	1.6026447203128\\
41	1.6026447203128\\
42	1.6026447203128\\
43	1.6026447203128\\
44	1.6026447203128\\
45	1.6026447203128\\
46	1.6026447203128\\
47	1.6026447203128\\
48	1.6026447203128\\
49	1.6026447203128\\
50	1.6026447203128\\
};
\addlegendentry{\sffamily{Zero-Forcing \& Final projection \cite{ca}}}

\end{axis}
\end{tikzpicture}%
    \caption{Performance results of the sum rate for the analog-only architecture: ZF-Max-Power considered for different approaches and for various number of ZF cyclic maximization (outer iterations).}
    \label{fig2}
\end{figure}

\textcolor{black}{\subsection{Convergence}
Fig.~\ref{fig2} presents achievable rates with respect to ZF cycle maximization. The cost function in \cite{ca} converges after one iteration whereas the all-digital ZF solution \cite{unconst} and our proposed hybrid beamforming approach require roughly five iterations each. Consequently, this confirms the convergence of the cost function of our approach toward the optimal all-digital and CA-constrained solutions. In \cite{ca}, the authors proposed a two-stage beamforming design. The first stage is based on the ZF projection whereas the second projects the beamformers onto the CA subspace, separately from the ZF projection. This methods yield low spectral efficiency of around 2 bits/s/Hz whereas the proposed method offers approximately 12 bits/s/Hz. In fact, the second projection onto the CA subspace in \cite{ca}, which is handled separately from the first ZF projection, eventually violates the ZF constraints.  The constraint violation incurs huge SI power and hence substantial degradation in spectral efficiency. In our proposed method, the two projections are performed jointly using alternating projections to avoid the constraint violations and iterate towards the optimal solution.}

\subsection{Proposed vs.\ Conventional Approaches}

\begin{figure}[tb]
\centering
\setlength\fheight{5.5cm}
\setlength\fwidth{7.5cm}
\input{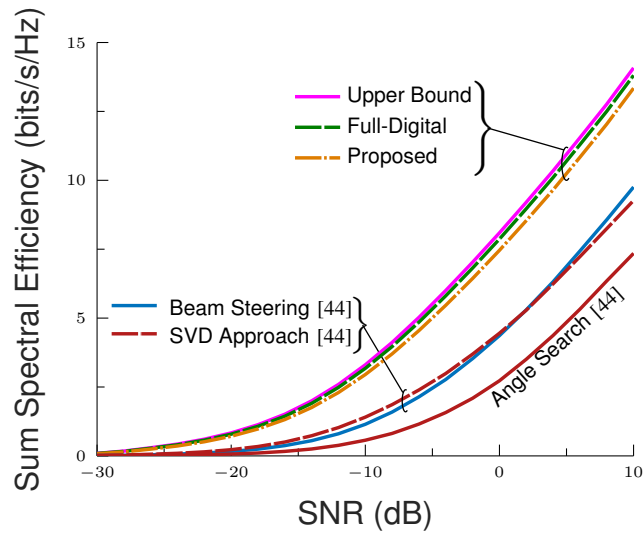}
    \caption{Sum spectral efficiency results: Comparisons are made between the proposed and conventional approaches. The conventional techniques in \cite{unconst} are developed for machine-to-machine FD systems. In this work, we expanded these techniques accordingly to support the proposed wideband system model. The performance is evaluated for an analog-only architecture in accordance with \cite{unconst} in the sub-6 GHz band. }
    \label{pict4}
\end{figure}
Fig. \ref{pict4} compares proposed and conventional approaches implemented in analog-only architectures. We observe that conventional designs are very sensitive to the SI because the zero-forcing and phase shift constraints are not properly handled. In contrast, the proposed design is resilient to SI and achieves higher sum spectral efficiency of around 13 bits/s/Hz at 10 dB of SNR whereas beam steering, SVD and angle search techniques achieve roughly 9, 9, and 7 bits/s/Hz.

\subsection{Degrees of Freedom}
\begin{figure}[tb]
\centering
\setlength\fheight{5.5cm}
\setlength\fwidth{7.5cm}
\input{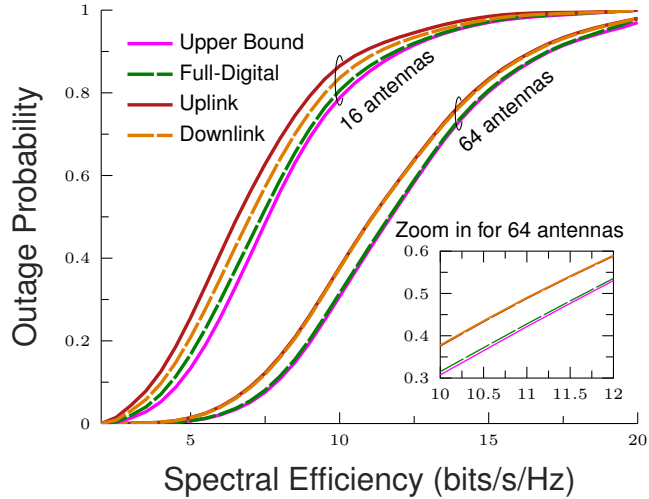}
    \caption{Outage probability results: Comparisons are made between uplink and downlink scenarios for hybrid beamforming with full-digital as well as upper bound which serve as benchmarking tools. In addition, the comparison involves two settings with 16 and 64 antennas at the BS/relay. MmWave configuration is considered in this scenario while the SNR is maintained at 0 dB while $N_{\text{inner}}=50$ iterations. }
    \label{pict6}
\end{figure}
Fig. \ref{pict6} illustrates outage probability vs. spectral efficiency for 16 and 64 antennas at the relay/BS. For the same number of inner alternating projection iterations, the gap between uplink and downlink rates for hybrid architectures changes with the number of antennas. As we conclude so far, the rate performance improves with the number of iterations of alternating projections; however, this improvement is also limited by the number of antennas. The antennas are shared between providing spatial multiplexing gain (sustaining enough spatial streams) and SI cancellation. If the number of antennas is too low, there is not enough DoF to eliminate the SI, and hence, the alternating projections routine becomes useless because increasing $N_{\text{inner}}$ does not bring any further improvement. The 64-antenna setting, which has enough degrees of freedom, provides not only the spectral efficiency (multiplexing gain) but also perfectly cancels the SI which is observed by the complete match between the uplink and downlink rates. 

\subsection{Power Consumption and RF Chains}
\begin{figure}
\centering
\setlength\fheight{5.5cm}
\setlength\fwidth{7.5cm}
%
%

\definecolor{mycolor1}{rgb}{1.00000,0.00000,1.00000}%
\definecolor{mycolor2}{rgb}{0.00000,0.49804,0.00000}%
\definecolor{mycolor3}{rgb}{0.00000,0.44706,0.74118}%
\definecolor{mycolor4}{rgb}{0.87059,0.49020,0.00000}%
\definecolor{cornellred}{rgb}{0.7, 0.11, 0.11}
\begin{tikzpicture}

\begin{axis}[%
width=0.951\fwidth,
height=\fheight,
at={(0\fwidth,0\fheight)},
scale only axis,
xmin=0,
xmax=14,
xlabel style={font=\color{white!15!black}},
xlabel={\textsf{Number of RF Chains}},
ymin=0,
ymax=6,
ylabel style={font=\color{white!15!black}},
ylabel={\textsf{Energy Efficiency (bits/Joule/Hz)}},
axis background/.style={fill=white},
axis x line*=bottom,
axis y line*=left,
legend style={legend cell align=left, align=left, draw=none,fill=none}
]
\addplot [color=mycolor1, line width=1.3pt]
  table[row sep=crcr]{%
1	4.40577356349385\\
2	5.52086572332713\\
3	5.50687584487464\\
4	5.03681193848618\\
5	4.18983928053848\\
6	3.05568706690332\\
7	2.06331440521361\\
8	1.29510753033586\\
9	0.771476863660851\\
10	0.441345517699016\\
11	0.235598651002849\\
12	0.124037127699909\\
13	0.061599562807192\\
14	0.0229163013698764\\
};
\addlegendentry{\textsf{Variant 1}}

\addplot [color=mycolor2, dash pattern={on 10pt off 1pt on 0pt off 0pt}, line width=1.3pt]
  table[row sep=crcr]{%
1	2.61255720132411\\
2	3.60011469252724\\
3	3.88559394592794\\
4	3.74707512412952\\
5	3.54182475216733\\
6	3.27891040220639\\
7	3.04939910297844\\
8	2.72675095373126\\
9	2.41588002083983\\
10	2.1347069483329\\
11	1.83719141707841\\
12	1.53922889975128\\
13	1.23689975114746\\
14	1.02395862566205\\
};
\addlegendentry{\textsf{Variant 2}}

\addplot [color=mycolor3, dash pattern={on 10pt off 1pt on 1pt off 1pt}, line width=1.3pt]
  table[row sep=crcr]{%
1	2.5330332029839\\
2	3.38395373423599\\
3	3.5765545915764\\
4	3.55033434845953\\
5	3.35055012359813\\
6	3.06940876503319\\
7	2.7794691453853\\
8	2.50164450304218\\
9	2.22320762999804\\
10	1.90673765678991\\
11	1.67079809791451\\
12	1.39951045049681\\
13	1.17197758529628\\
14	0.9526463643527\\
};
\addlegendentry{\textsf{Variant 3}}

\addplot [color=mycolor4, dotted, line width=1.3pt]
  table[row sep=crcr]{%
1	1.58235599179846\\
2	2.29965538636777\\
3	2.61145416607309\\
4	2.7104871782125\\
5	2.63279188818791\\
6	2.55166629706988\\
7	2.43410095942176\\
8	2.2091627823809\\
9	2.08415038372318\\
10	1.9326146349194\\
11	1.79548901026595\\
12	1.69988116960104\\
13	1.58983054814464\\
14	1.45347361163997\\
};
\addlegendentry{\textsf{Variant 4}}

\end{axis}
\end{tikzpicture}%
    \caption{Energy efficiency performance: Results are evaluated for four variants and for different number of RF chains at the relay/BS. The SNR is fixed at 0 dB. The simulations are performed for uplink scenario with $N_{\text{inner}}=50$ iterations.}
    \label{pict7}
\end{figure}
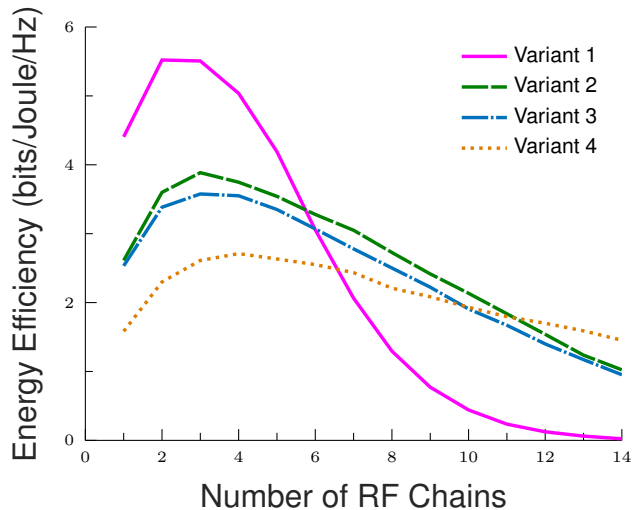

Next, we evaluate energy efficiency for four variants over three different mmWave bands per Table \ref{setting}.
Fig. \ref{pict7} shows energy efficiency vs. number of RF chains at the relay/BS. For a small number of RF chains, variant 1 provides the best energy efficiency. For a smaller number of RF chains, the rate at which the power consumption increases is higher that the rate that spectral efficiency increases. As the number of RF chains increases, however, the spectral efficiency improves at a higher rate; variant 4 has the best energy efficiency because it provides hefty beamforming gain from the massive number of antennas. The effect of the ADC power can be observed by comparing variants 2 and 3 (which have same number of antennas operating in the same band). Variant 2 is more power-efficient than variant 3 since the latter requires more ADC power. Also, variant 2 is more power-efficient than variant 3 vs. number of RF chains.

\begin{table}[tb]
\renewcommand{\arraystretch}{1}
\caption{Parameters of Each Variant \cite{ni,heath}.}
\label{setting}
\centering
\small
\begin{tabular}{rllll}
\bfseries Parameter & \bfseries Variant 1 & \bfseries Variant 2 & \bfseries Variant 3 & \bfseries Variant 4\\
\hline
 Frequency & 28 GHz & 39 GHz & 39 GHz & 73 GHz\\
 Bandwidth & 850 MHz & 1.4 GHz & 1.6 GHz & 2 GHz\\
 ADC Power & 250 mW & 400 mW & 450 mW & 550 mW\\
BS Antennas & 32 & 64 & 64 & 128\\
\end{tabular}
\end{table}

\begin{figure}[tb]
\centering
\setlength\fheight{5.5cm}
\setlength\fwidth{7.5cm}
%
%
\definecolor{mycolor1}{rgb}{1.00000,0.00000,1.00000}%
\definecolor{mycolor2}{rgb}{0.00000,0.49804,0.00000}%
\definecolor{mycolor3}{rgb}{0.00000,0.44706,0.74118}%
\definecolor{mycolor4}{rgb}{0.87059,0.49020,0.00000}%
\definecolor{cornellred}{rgb}{0.7, 0.11, 0.11}
\begin{tikzpicture}

\begin{axis}[%
width=0.951\fwidth,
height=\fheight,
at={(0\fwidth,0\fheight)},
scale only axis,
xmin=10,
xmax=70,
xlabel style={font=\color{white!15!black}},
xlabel={\sffamily{Number of BS Antennas}},
ymin=-20,
ymax=100,
ylabel style={font=\color{white!15!black}},
ylabel={\sffamily{Uplink Spectral Efficiency Gain [\%]}},
axis background/.style={fill=white},
axis x line*=bottom,
axis y line*=left,
legend style={at={(0.97,0.03)}, anchor=south east, legend cell align=left, align=left, draw=none}
]
\addplot [color=mycolor1, line width=1.3pt]
  table[row sep=crcr]{%
10	43.5836187443942\\
12	55.917089331784\\
14	65.0398184545693\\
16	73.6206898144156\\
18	80.0383579136289\\
20	85.1430627099481\\
22	88.7933574764564\\
24	91.4862237599041\\
26	93.4225295111141\\
28	94.8864258427508\\
30	95.9559674750223\\
32	96.7874573586765\\
34	97.3779859240217\\
36	97.8722872507347\\
38	98.2747434653278\\
40	98.5827089428078\\
42	98.8091202672966\\
44	99.027615196959\\
46	99.1838762010991\\
48	99.2973421196621\\
50	99.3908738348265\\
52	99.4834192970618\\
54	99.5186456868662\\
56	99.5545011651053\\
58	99.5886063804403\\
60	99.619330135326\\
62	99.6390147761102\\
64	99.6841345758463\\
66	99.6707427514113\\
68	99.6652562429298\\
70	99.5346342229615\\
};
\addlegendentry{\sffamily{HD (10 Outer, 20 Inner)}}

\addplot [color=mycolor2,dash pattern={on 10pt off 1pt on 0pt off 0pt} ,line width=1.3pt]
  table[row sep=crcr]{%
10	39.6347525373357\\
12	52.9995289196009\\
14	62.8644391149264\\
16	72.1325316783492\\
18	79.0808952638004\\
20	84.4424011497351\\
22	88.2892694174711\\
24	91.0011705603451\\
26	92.9486792416275\\
28	94.4108729241696\\
30	95.6012710129911\\
32	96.5201516733301\\
34	97.2312835671857\\
36	97.8049287687887\\
38	98.2204118457919\\
40	98.5197649165968\\
42	98.762778797043\\
44	98.9826790842427\\
46	99.1761589850053\\
48	99.3569505603702\\
50	99.4867940697429\\
52	99.575609948\\
54	99.6415328727123\\
56	99.6678595442687\\
58	99.6817539146428\\
60	99.7149174291794\\
62	99.7543589588575\\
64	99.7964537924434\\
66	99.828944879303\\
68	99.8686461246726\\
70	99.8424730647688\\
};
\addlegendentry{\sffamily{HD (3 Outer, 20 Inner)}}

\addplot [color=mycolor3,dash pattern={on 10pt off 1pt on 1pt off 1pt}, line width=1.3pt]
  table[row sep=crcr]{%
10	-14.5958570682468\\
12	3.01982375731831\\
14	17.1218322816921\\
16	30.5889258656795\\
18	41.6223178197007\\
20	50.9825165976006\\
22	58.5457458714127\\
24	64.8015550068156\\
26	69.9128666297818\\
28	74.2085481063799\\
30	77.7511857989293\\
32	80.7347496768853\\
34	83.1925202163594\\
36	85.303391910564\\
38	87.0396729654374\\
40	88.4323480955608\\
42	89.5898463010026\\
44	90.5552661016798\\
46	91.319236453638\\
48	92.0268206181069\\
50	92.7439238138286\\
52	93.3851276611399\\
54	93.9680299216413\\
56	94.4890610522663\\
58	94.893197110376\\
60	95.2513449102322\\
62	95.5629260523792\\
64	95.8564639548223\\
66	96.036876701029\\
68	96.2632895479445\\
70	96.177598581493\\
};
\addlegendentry{\sffamily{HD (3 Outer,15 Inner)}}

\addplot [color=cornellred, line width=1.3pt]
  table[row sep=crcr]{%
10	37.7605539123173\\
12	50.5612673229636\\
14	59.7604240058537\\
16	68.4539048134773\\
18	74.8309742254493\\
20	79.97160040961\\
22	83.6052658692861\\
24	86.2601838081522\\
26	88.1751415671986\\
28	89.6421022665835\\
30	90.6975240553759\\
32	91.5315187851635\\
34	92.1657690113876\\
36	92.6944493355957\\
38	93.0514576990913\\
40	93.3445329197953\\
42	93.5705815876755\\
44	93.7573133956038\\
46	93.8862879951917\\
48	94.0244452236943\\
50	94.1423513703117\\
52	94.2463164981112\\
54	94.3265880630116\\
56	94.4042171054725\\
58	94.4522615206422\\
60	94.4734129481601\\
62	94.5025145509839\\
64	94.5524412693456\\
66	94.6299704092216\\
68	94.7164998526495\\
70	94.8668122763326\\
};
\addlegendentry{\sffamily{SVD (10 Outer, 20 Inner)}}

\addplot [color=mycolor4,dash pattern={on 10pt off 1pt on 0pt off 0pt}, line width=1.3pt]
  table[row sep=crcr]{%
10	32.4247968206317\\
12	46.2510895493928\\
14	56.1558776753063\\
16	65.5499924429411\\
18	72.4512117273179\\
20	77.9134063431364\\
22	81.8006563523889\\
24	84.7183581649707\\
26	86.7971404115261\\
28	88.3779654467768\\
30	89.5428888054065\\
32	90.4345407318711\\
34	91.133370970836\\
36	91.6933341341088\\
38	92.1763402423549\\
40	92.5454958449514\\
42	92.8485012194724\\
44	93.0708955574673\\
46	93.2438275836916\\
48	93.3405544668809\\
50	93.4285646053856\\
52	93.4962575194262\\
54	93.5516965480086\\
56	93.5926513718097\\
58	93.6348063654263\\
60	93.6727405822097\\
62	93.7150199864948\\
64	93.7810781504221\\
66	93.8552663506876\\
68	93.939036246943\\
70	94.0294729594005\\
};
\addlegendentry{\sffamily{SVD (3 Outer, 20 Inner)}}

\addplot [color=black, dash pattern={on 10pt off 1pt on 1pt off 1pt},line width=1.3pt]
  table[row sep=crcr]{%
10	-16.2974349444492\\
12	-0.31115688429149\\
14	12.8302470910862\\
16	25.356305154922\\
18	35.828882687209\\
20	44.8457861269016\\
22	52.2511976212024\\
24	58.4412978593982\\
26	63.6186982677155\\
28	67.9607304216208\\
30	71.6228245540272\\
32	74.6027890457755\\
34	76.948951408129\\
36	78.883313620885\\
38	80.4857908543945\\
40	81.8690521639036\\
42	83.1320394133516\\
44	84.3047238677466\\
46	85.3289310685401\\
48	86.1687596268871\\
50	86.8729271717897\\
52	87.4601209100382\\
54	87.9215297513037\\
56	88.3121145843039\\
58	88.7026267255278\\
60	89.0215086976733\\
62	89.3070751365756\\
64	89.655920548154\\
66	90.0443519245437\\
68	90.3980384138333\\
70	90.9077855310136\\
};
\addlegendentry{\sffamily{SVD (3 Outer, 15 Inner)}}

\end{axis}
\end{tikzpicture}%
    \caption{Performance results: uplink achievable rate gain of FD relative to SVD and HD mode for different number of BS antennas, number of inner and outer iterations.}
     \label{fig6}
\end{figure}
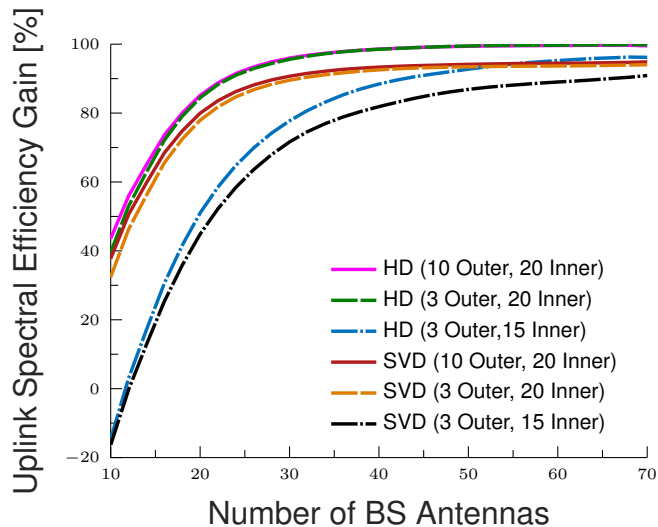

\textcolor{black}{\subsection{Spectral Efficiency Gain}
Fig.~\ref{fig6} presents the FD rate gain relative to SVD and HD vs. number of BS antennas for different numbers of outer and inner iterations. A goal is to achieve acceptable gain relative to HD. With a small number of antenna (10), FD achieves a gain of roughly 42\% vs. HD and 34\% vs. SVD. As the number of inner iterations decreases, the alternating projections algorithm will not converge to the optimal beamformers and will perform worse than SVD or HD due to leftover SI power and the CA constraint. As the number of antennas increases, enough DoF become available to suppress SI and outperform HD and SVD. This gain further improves by increasing outer and inner iterations to obtain convergence and hence arrive at the best beamformers to maximize the achievable rate. The proposed design scales better with the number of antennas as the gain converges to approximately 98.7\% and 94\% vs. HD and SVD, respectively. The system could converge with fewer iterations but more antennas would be needed.}

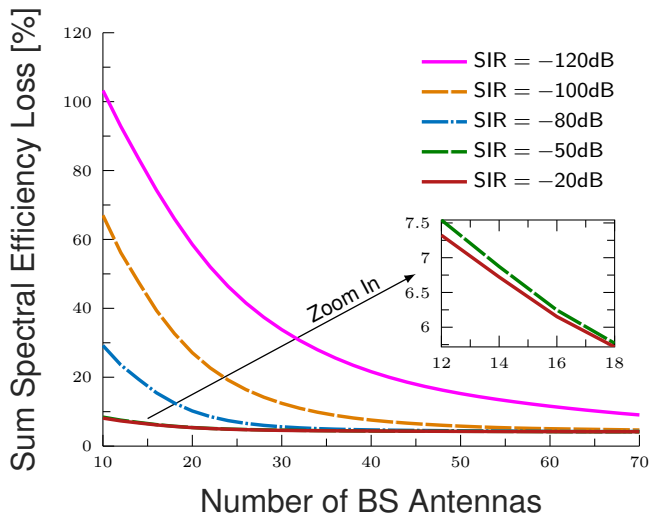
\begin{figure}[tb]
\centering
\setlength\fheight{5.5cm}
\setlength\fwidth{7.5cm}
%
%
\definecolor{mycolor1}{rgb}{1.00000,0.00000,1.00000}%
\definecolor{mycolor2}{rgb}{0.00000,0.49804,0.00000}%
\definecolor{mycolor3}{rgb}{0.00000,0.44706,0.74118}%
\definecolor{mycolor4}{rgb}{0.87059,0.49020,0.00000}%
\definecolor{cornellred}{rgb}{0.7, 0.11, 0.11}
\begin{tikzpicture}

\begin{axis}[%
width=0.951\fwidth,
height=\fheight,
at={(0\fwidth,0\fheight)},
scale only axis,
xmin=10,
xmax=70,
xlabel style={font=\color{white!15!black}},
xlabel={\sffamily{Number of BS Antennas}},
ymin=0,
ymax=120,
ylabel style={font=\color{white!15!black}},
ylabel={\sffamily{Sum Spectral Efficiency Loss [\%]}},
axis background/.style={fill=white},
axis x line*=bottom,
axis y line*=left,
legend style={legend cell align=left, align=left, draw=none}
]

\draw [-latex,black,line width=.5pt] (15,8) to (45,50);
\node[right, align=left, rotate=28]
at (axis cs:32,36) {\scriptsize{\sffamily{Zoom In}}};

\addplot [color=mycolor1, line width=1.3pt]
  table[row sep=crcr]{%
10	103.223959036193\\
12	92.8238429389054\\
14	83.4012111989138\\
16	74.2691318225605\\
18	65.9945940050882\\
20	58.5474780984562\\
22	52.0303850511645\\
24	46.3770852334211\\
26	41.5382026219781\\
28	37.3903856847249\\
30	33.8102885413261\\
32	30.6915551598812\\
34	27.9690910276221\\
36	25.5717590448561\\
38	23.4579077168645\\
40	21.6108279225864\\
42	19.9960192053077\\
44	18.5748968560443\\
46	17.329363307816\\
48	16.2337493761335\\
50	15.2569103077067\\
52	14.3822914558589\\
54	13.5930073492033\\
56	12.8693661827033\\
58	12.2001674391064\\
60	11.5773669273439\\
62	10.9995540324928\\
64	10.4614542685351\\
66	9.96569647168636\\
68	9.48522283012701\\
70	9.05782662549609\\
};
\addlegendentry{$\mathsf{SIR = -120 dB}$}

\addplot [color=mycolor4,dash pattern={on 10pt off 1pt on 0pt off 0pt} ,line width=1.3pt]
  table[row sep=crcr]{%
10	67.0005157364482\\
12	56.1699419963774\\
14	47.5548974117228\\
16	39.3479239319013\\
18	32.6672197651205\\
20	27.1004145687558\\
22	22.6999622914203\\
24	19.1809089544395\\
26	16.4052558679424\\
28	14.196861041367\\
30	12.4247781223835\\
32	10.9909126161077\\
34	9.84218894585158\\
36	8.91596901230172\\
38	8.16844280526189\\
40	7.56295463211229\\
42	7.07596545070646\\
44	6.67376489377776\\
46	6.3354285876582\\
48	6.04492062409117\\
50	5.79849987278871\\
52	5.58673827858959\\
54	5.40761989583542\\
56	5.26341356491746\\
58	5.14966437585997\\
60	5.05573897925125\\
62	4.9763166844893\\
64	4.90374255382896\\
66	4.83101289397091\\
68	4.75792605477268\\
70	4.67992175262797\\
};
\addlegendentry{$\mathsf{SIR = -100 dB}$}

\addplot [color=mycolor3,dash pattern={on 10pt off 1pt on 1pt off 1pt}, line width=1.3pt]
  table[row sep=crcr]{%
10	29.2133768030354\\
12	23.5683521856106\\
14	19.3907839129697\\
16	15.456127954348\\
18	12.4964172280592\\
20	10.2009496604597\\
22	8.5702890146941\\
24	7.40466205067861\\
26	6.59972122482299\\
28	6.02389901201982\\
30	5.60982979629991\\
32	5.31060831329153\\
34	5.08807210671684\\
36	4.92257073699108\\
38	4.80221480463035\\
40	4.71014650980265\\
42	4.63802180709035\\
44	4.58388010630522\\
46	4.53762056461105\\
48	4.49486357544275\\
50	4.45604193681784\\
52	4.41921426172916\\
54	4.38131194449074\\
56	4.34532576671267\\
58	4.31176027548565\\
60	4.27994405443971\\
62	4.25516409876887\\
64	4.23746955269021\\
66	4.23786063360912\\
68	4.23996495419294\\
70	4.27861283925346\\
};
\addlegendentry{$\mathsf{SIR = -80 dB}$}

\addplot [color=mycolor2, dash pattern={on 10pt off 1pt on 0pt off 0pt},line width=1.3pt]
  table[row sep=crcr]{%
10	8.42490870345\\
12	7.54228658914079\\
14	6.8728420792984\\
16	6.2462346451795\\
18	5.7690548556246\\
20	5.40226639137499\\
22	5.14181508679189\\
24	4.95490751107037\\
26	4.82297862453445\\
28	4.72151907206967\\
30	4.63966491184272\\
32	4.57155864540476\\
34	4.5136282744324\\
36	4.46235517718686\\
38	4.41989138985971\\
40	4.38440741670772\\
42	4.35646024819725\\
44	4.33670656427969\\
46	4.32122178291556\\
48	4.30994166476567\\
50	4.29932565761571\\
52	4.28505336358736\\
54	4.26489330849776\\
56	4.2442553176743\\
58	4.22001394688107\\
60	4.19665674479587\\
62	4.17824877798654\\
64	4.16665828983828\\
66	4.15962212316708\\
68	4.15659389254379\\
70	4.16006691965425\\
};
\addlegendentry{$\mathsf{SIR = -50 dB}$}

\addplot [color=cornellred, line width=1.3pt]
  table[row sep=crcr]{%
10	8.12933770185673\\
12	7.32326944567631\\
14	6.72184795327066\\
16	6.15275689841053\\
18	5.71687671620598\\
20	5.36635406071651\\
22	5.10661180722658\\
24	4.91041201978445\\
26	4.76216939772323\\
28	4.65022216039131\\
30	4.56859090940296\\
32	4.50833079900179\\
34	4.46318910843083\\
36	4.43340624659061\\
38	4.41115902513902\\
40	4.38920228754815\\
42	4.36667213275348\\
44	4.34480957938152\\
46	4.32126962154278\\
48	4.29717844832937\\
50	4.27732461447668\\
52	4.26049059260446\\
54	4.24476436002884\\
56	4.22945705765975\\
58	4.21676609587852\\
60	4.20397806923143\\
62	4.19285784572532\\
64	4.18420155747883\\
66	4.17887047823209\\
68	4.17397605291471\\
70	4.17286331077599\\
};
\addlegendentry{$\mathsf{SIR = -20 dB}$}
\end{axis}

\begin{axis}[%
width=0.307\fwidth,
height=0.307\fheight,
at={(0.6\fwidth,0.24\fheight)},
scale only axis,
xmin=12,
xmax=18,
ymin=5.71687671620598,
ymax=7.54228658914079,
axis background/.style={fill=white},
legend style={legend cell align=left, align=left, draw=white!15!black}
]

\addplot [color=mycolor2, dash pattern={on 10pt off 1pt on 0pt off 0pt},line width=1.3pt]
  table[row sep=crcr]{%
12	7.54228658914079\\
14	6.8728420792984\\
16	6.2462346451795\\
18	5.7690548556246\\
};

\addplot [color=cornellred, line width=1.3pt]
  table[row sep=crcr]{%
12	7.32326944567631\\
14	6.72184795327066\\
16	6.15275689841053\\
18	5.71687671620598\\
};

\end{axis}
\end{tikzpicture}%
    \caption{Performance results for the average throughput loss of the proposed hybrid beamforming algorithm relative to the upper bound (without SI) for different values of the SIR. 20 inner iterations and 10 outer iterations considered for this simulations.}
     \label{figloss}
\end{figure}

\textcolor{black}{\subsection{Spectral Efficiency Loss}
Fig.~\ref{figloss} illustrates average rate loss of the proposed hybrid beamforming algorithm relative to the upper bound vs. number of BS antennas and SIR values. For 10 BS antennas and -120 dB of SIR for a cell-edge user, the average rate loss is nearly 100\%. As the number of BS antennas increases, the average rate loss saturates at 10\% and the loss decreases to approximately 5\% for the remaining SIR values. As the number of antennas becomes massive, the rate loss for -120 dB of SIR is expected to converge to the 5\% threshold since the BS will have more than enough DoF for SI cancellation.
Overall, the average rate loss is acceptable for 64 BS antennas, even for -120 dB of SIR wherein the loss is roughly 15\%. Thereby, for a reasonable number of BS antennas, the proposed technique reduces the average throughput loss to 5\% for middle and near users. This 5\% residual loss cannot be avoided because it is incurred from the CA constraint.} 

\section{Conclusion}
\textcolor{black}{For mmWave communications using hybrid analog/digital beamformers, we analyzed a single-cell full-duplex system. Self-interference at the basestation is due to the basestation transmitting to a downlink UE and receiving from an uplink UE simultaneously using the same resource blocks. We propose iterative algorithms to design full-digital and hybrid beamformers to maximize uplink plus downlink sum capacity while bringing self-interference below the noise floor. Since we use the analog beamformers built into 5G mmWave systems, our approach does not need additional circuitry. In the hybrid beamformer design, a key insight was to move zero forcing to the analog domain to help prevent ADC saturation. In simulation, the proposed algorithms improved spectral efficiency vs. inner alternating projection iterations; overall enhancement is limited by the degrees of freedom. Our algorithms showed higher spectral efficiency than \textcolor{black}{half-duplex systems} as well as full-duplex systems using beamsteering, SVD, and angle search. We also evaluated energy efficiency in the 28, 39 and 73 GHz mMWave bands.}

\appendices
\textcolor{black}{\begin{center}{\sc Appendix A: Convergence of Alternating Projections Method}\end{center}
We prove the convergence of the alternating projections method for the case $A\cap B \neq \emptyset$. The proof of convergence is similar to the proof when $A\cap B = \emptyset$ \cite{Cheney}. 
Let $\bar{x}$ be any point in the intersection $A\cap B$. We claim that each projection brings the point closer to $\bar{x}$. To see this, we first observe that since $y_k$ is the projection of $x_k$ onto $B$. we have
\begin{equation}
    B \subseteq \{ z| \left( x_k-y_k \right)^T (z-y_k)\leq 0  \}.
\end{equation}
In other words, the halfspace passing through $y_k$, with outward normal $x_k-y_k$, contains $B$. This follows from the optimality conditions for Euclidean projection, and can be shown directly: if any point of $B$ were on the other side of the hyperplane, a small step from $y_k$ towards the point would give a point in $B$ that is closer to $x_k$ than $y_k$, which is impossible.
Now we note that
\begin{equation}
\begin{split}
       \|x_k - \bar{x} \|^2 &= \|x_k-y_k+y_k-\bar{x} \|^2\\&= \|x_k-y_k  \|^2 + \|y_k - \bar{x}\|^2 + 2\left(x_k -y_k\right)^T\left( y_k-\bar{x} \right) \\&\geq \|x_k-y_k  \|^2 + \| y_k - \bar{x} \|^2
\end{split}
\end{equation}
Using the observation above. Thus we have 
\begin{equation}\label{eq1}
\| y_k - \bar{x}\|^2 \leq \| x_k-\bar{x}\|^2 - \|y_k-x_k\|^2    
\end{equation}
This shows that $y_k$ is closer to $\bar{x}$ than $x_k$ is. In a similar way, we can show that
\begin{equation}\label{eq2}
\|x_{k+1} - \bar{x}\|^2 \leq \|y_k - \bar{x}  \|^2 - \| x_{k+1} - y_k \|^2    
\end{equation}
i.e., $x_{k+1}$ is closer to $\bar{x}$ than $y_k$ is.
We observe that all iterates are no farther from $\bar{x}$ than $x_0$:
\begin{equation}\label{eq3}
    \|x_k-\bar{x}\| \leq \|x_0-\bar{x} \|,~~\|y_k - \bar{x} \| \leq \|x_0 - \bar{x}\|,~~k=1,2,\ldots
\end{equation}
We conclude that sequences $x_k$ and $y_k$ are bounded. Therefore, sequence $x_k$ has an accumulation point $x^*$. Since $B$ is closed, and $x_k\in A$, we have $x^*\in A$. Next, We show that $x^*\in B$, and both sequences $x_k$ and $y_k$ converge to $x^*$.
From (\ref{eq1}) and (\ref{eq2}), we find that the sequence
\begin{equation}
\| x_0 - \bar{x}\|,~\|y_0-\bar{x}\|,~\|x_1 - \bar{x} \|,~\|y_1 - \bar{x}\|,\ldots    
\end{equation}
is decreasing, and so it converges. We then conclude from (\ref{eq1}) and (\ref{eq2}) that $\|y_k-x_k \|$ and $\|x_{k+1}-y_k\|$ must converge to zero.
A subsequence of $x_k$ converges to $x^*$. From 
\begin{equation}
\mathbf{dist}\left(x_k,B\right) = \mathbf{dist}\left(x_k,y_k\right)\xrightarrow{}0    
\end{equation}
and closedness of $B$, $x^*\in B$. Thus, $x^*\in A\cap B$.
Since $x^*$ is in the intersection, we take $\bar{x}=x^*$ above (since $\bar{x}$ is any point in the intersection). The distance of $x_k$ and $y_k$ to $x^*$ is decreasing. Since a subsequence converges to zero, we conclude $\|x_k-x^*\|$ and $\|y_k-x^*\|$ converge to zero.}

\vspace*{-0.2in}
\textcolor{black}{\begin{center}{\sc Appendix B: Full-Digital and Analog Solutions}\end{center}
Consider a matrix $\textbf{C}$ with full rank $P$ (if not, it can be replaced in the constraint by another matrix $\textbf{C}^{'} \in \mathbb{C}^{M\times P^{'}}$ with the same range as \textbf{C} and of full rank $P^{'}<P$). Let the columns of $\textbf{U}_0 \in \mathbb{C}^{M\times (M-P)}$ constitute an orthogonal basis for the subspace orthogonal to \textbf{C}. Then $\textbf{P}_{\perp} = \textbf{U}_0\textbf{U}_0^*$ and $\textbf{X}^*\textbf{C} = \textbf{0}$ implies that $\textbf{X} = \textbf{U}_0\textbf{Y}$ for some $\textbf{Y} \in \mathbb{C}^{(M-P)\times N}$. Further, $\textbf{X}^*\textbf{X} = \textbf{I}_N$ and $M-P \geq N$ imply that $\textbf{Y}^*\textbf{Y} = \textbf{I}_N$, since $\textbf{U}_0^*\textbf{U}_0 = \textbf{I}_{M-P}$. The problem becomes
\begin{equation}\label{opt1}
\begin{split}
 \max\limits_{\textbf{Y}^*\textbf{Y}=\textbf{I}_N} \log\det\left( \textbf{I}_N + \rho \textbf{Y}^*\textbf{U}_0^*\textbf{A}\textbf{A}^*\textbf{U}_0\textbf{Y} \right)
\end{split}
\end{equation}
whose solution $\textbf{Y}_{\star}$ is the $N$ dominant left singular vectors of $\textbf{U}_0^*\textbf{A}$. Thus, $\textbf{U}_0^*\textbf{A}$ admits an SVD of the form $\textbf{U}_0^*\textbf{A} = \textbf{Y}_{\star}\textbf{S}\textbf{V}^*$; i.e., $\textbf{U}_0\textbf{U}_0^*\textbf{A} = \textbf{U}_0\textbf{Y}_{\star}\textbf{S}\textbf{V}^*=\textbf{X}_{\star}\textbf{S}\textbf{V}^*$ constitutes an SVD of $\textbf{U}_0\textbf{U}_0^*\textbf{A}=\textbf{P}_{\perp}\textbf{A}$. Hence, the solution $\textbf{X}_{\star} = \textbf{U}_0\textbf{Y}_{\star}$ is given by the $N$ dominant left singular vectors of $\textbf{P}_{\perp}\textbf{A}$.} 

\vspace*{-0.2in}
\textcolor{black}{\begin{center}{\sc Appendix C: Digital Solution}\end{center}
For $\textbf{X}_{\text{RF}}$ given, consider the SVD $\textbf{X}_{\text{RF}} = \textbf{U}_{\text{RF}}\textbf{S}_{\text{RF}}\textbf{V}_{\text{RF}}^*$, and let $\textbf{Q} = \textbf{S}_{\text{RF}}\textbf{V}_{\text{RF}}^*\textbf{X}_{\text{BB}} \in \mathbb{C}^{L\times N}$. Then $\textbf{X}_{\text{RF}}\textbf{X}_{\text{BB}}=\textbf{U}_{\text{RF}}\textbf{Q}$ so that $\textbf{X}_{\text{BB}}^*\textbf{X}_{\text{RF}}^*\textbf{X}_{\text{RF}}\textbf{X}_{\text{BB}}=\textbf{Q}^*\textbf{U}_{\text{RF}}^*\textbf{U}_{\text{RF}}\textbf{Q}=\textbf{Q}^*\textbf{Q}$. Problem in terms of $\textbf{Q}$ is
\begin{equation}\label{opt2}
\begin{split}
 \max\limits_{\textbf{Q}^*\textbf{Q}=\textbf{I}_N} \log\det\left( \textbf{I}_N + \rho \textbf{Q}^*\textbf{U}_{\text{RF}}^*\textbf{A}\textbf{A}^*\textbf{U}_{\text{RF}}\textbf{Q} \right)
\end{split}
\end{equation}
whose solution $\textbf{Q}_{\star}$ is given by the $N$ dominant left singular vectors of $\textbf{U}_{\text{RF}}\textbf{A}$. By changing variables, we solve the problem $\textbf{X}_{\text{BB}}=\textbf{V}_{\text{RF}}\textbf{S}_{\text{RF}}^{-1}\textbf{Q}_{\star}$. For $\textbf{Q}=\textbf{Q}_{\star}$, the objective function becomes
\begin{equation}\label{opt3}
\begin{split}
\log\det\left( \textbf{I}_N + \rho \textbf{Q}^*\textbf{U}_{\text{RF}}^*\textbf{A}\textbf{A}^*\textbf{U}_{\text{RF}}\textbf{Q} \right) \leq \log\det\left( \textbf{I}_N + \rho \textbf{A}^*\textbf{A}\right)
\end{split}
\end{equation}
With the bound in (\ref{opt3}) applying to any semi-unitary $\textbf{U}_{\text{RF}}$. This bound holds with equality if the columns of $\textbf{U}_{\text{RF}}$ are taken as the $L$ dominant left singular vectors of \textbf{A}}.

\bibliographystyle{IEEEtran}
\bibliography{main}
\end{document}